\documentclass[useAMS,usenatbib]{mn2e}
\usepackage{times}
\usepackage{graphicx,epsfig} 
 
\title[Temperature and abundance profiles of groups] {Temperature and
  abundance profiles of hot gas in galaxy groups --\\ I. Results and
  statistical analysis}

\author[J. Rasmussen and T. J. Ponman]{Jesper Rasmussen\thanks{E-mail: 
    jesper@star.sr.bham.ac.uk} and Trevor J. Ponman \\ 
  School of Physics and Astronomy, University of Birmingham, 
  Edgbaston, Birmingham B15 2TT} 
\begin{document} 
 
\date{} 
 
\pagerange{\pageref{firstpage}--\pageref{lastpage}} \pubyear{2007} 
 
\maketitle 
 
\label{firstpage} 
 
\begin{abstract}
  The distribution of metals in groups of galaxies holds important
  information about the chemical enrichment history of the Universe.
  Here we present radial profiles of temperature and the abundance of
  iron and silicon of the hot intragroup medium for a sample of 15
  nearby groups of galaxies observed by {\em Chandra}, selected for
  their regular X-ray morphology.  All but one group display a cool
  core, the size of which is found to correlate with the mean
  temperature of the group derived outside this core. When scaled to
  this mean temperature, the temperature profiles are remarkably
  similar, being analogous to those of more massive clusters at large
  radii but significantly flatter inwards of the temperature peak.
  The Fe abundance generally shows a central excess followed by a
  radial decline, reaching a typical value of $0.1$~Z$_\odot$ within
  $r_{500}$, a factor of two lower than corresponding results for
  clusters.  Si shows less systematic radial variation, on average
  displaying a less pronounced decline than Fe and showing evidence
  for a flattening at large radii.  Off-centre abundance peaks are
  seen both for Fe and Si in a number of groups with well-resolved
  cores. Derived abundance ratios indicate that supernovae type Ia are
  responsible for 80~per~cent of the Fe in the group core, but the
  type~II contribution increases with radius and completely dominates
  at $r_{500}$.  We present fitting formulae for the radial dependence
  of temperature and abundances, to facilitate comparison to results
  of numerical simulations of group formation and evolution.  In a
  companion paper, we discuss the implications of these results for
  feedback and enrichment in galaxy groups.
\end{abstract} 
 
 
\section{Introduction}

Hot X-ray emitting gas constitutes the dominant baryonic component in
groups and clusters of galaxies.  The metals in this intracluster
medium (ICM) are believed to originate mainly in material ejected from
group and cluster galaxies by supernovae \citep{arna92}, with a
smaller portion driven out by galaxies through galaxy--galaxy and
galaxy--ICM interactions (\citealt{doma06}, but see also
\citealt{moll07}). Besides galactic stellar winds, other
non-gravitational processes have also played a role in redistributing
the enriched gas, and more generally, in shaping the thermodynamic
properties of the ICM. These processes include radiative cooling
\citep*{sand06}, and heating and mixing of gas by active galactic
nuclei \citep*{cros05,mcna05,osul05,rebu06,roed07}. The spatial
distribution of metals, and to some extent the temperature structure
of the ICM, reflect the action of such processes, thus offering
insight into the importance and nature of the mechanisms, other than
gravity, that have established the properties of the ICM at the
present epoch.  Groups of galaxies are particularly interesting in
this context because such processes are expected to be relatively more
important in low-mass systems due to their shallower gravitational
potentials.

Iron and silicon comprise the major diagnostic elements for studies of
the ICM abundance distribution, as these elements give rise to the
most prominent lines in the soft X-ray spectrum of thermal plasmas,
and are among the elements for which there is a distinctively
different yield in type~Ia and type~II supernova explosions according
to standard supernova (SN) models (see, e.g., \citealt{baum05} and
references therein). Whilst iron is predominantly produced by SN~Ia,
silicon is more evenly mixed between the two SN types.  The ratio of
Si to Fe abundance therefore provides valuable information on the
relative importance of SN~Ia vs.\ SN~II in enriching the ICM.
   
Abundance and temperature profiles of massive clusters have received
considerable attention (e.g.,
\citealt{degr01,baum05,vikh05,piff05,dona06}).  It is now well
established that clusters with cool central regions typically exhibit
high central Fe abundances (e.g., \citealt{degr01,irwi01,degr04})
which can be attributed predominantly to SN~Ia enrichment by the
stellar population of the central bright cluster galaxy
\citep*{fino00,boeh04,degr04}.  Clusters with no evidence for strong
central cooling tend to display a more uniform Fe distribution.

At large radii ($r\ga 0.5r_{200}$, where $r_{200}$ is the radius
enclosing a mean density of 200~times the critical density), $Z_{\rm
  Fe}$ typically drops to a value $\sim 0.2$~Z$_\odot$
\citep{fino00,degr01,tamu04,degr04}.  As suggested by, for example,
\citet{fino00}, this could result from SN~II--dominated enrichment at
an early stage in the cluster formation via energetic
starburst winds. Such a picture receives support from the observation
that a substantial fraction of the ICM metals were already in place by
$z\sim 1$ \citep{tozz03,bale07}.

The situation in lower-mass systems is much less clear, despite their
relatively larger importance for the cosmic baryon and metal budgets.
Historically, this has reflected the difficulty in assembling large
samples of groups bright enough to allow detailed, spatially resolved
X-ray spectroscopy.  Earlier works based on {\em ROSAT} and {\em ASCA}
data typically featured only a handful of $kT\la 2$~keV groups, often
included as part of larger cluster-dominated samples (e.g.\
\citealt{fuka98,fino99,fino00}; \citealt*{fino01a}). Such studies
suggested a $Z_{\rm Si}/Z_{\rm Fe}$ ratio roughly solar in groups
\citep*{fuka96,fuka98,davi99,hwan99}, about half the value found in
early cluster studies \citep{mush96}, with an indication that Si/Fe
rises above the solar ratio outside group cores (e.g.\
\citealt{fuka98}; \citealt{fino99}).  However, the sensitivity and
spatial resolution of {\em ROSAT} and {\em ASCA} imposed natural
limits to the amount of detail one could obtain on the spatial
variation of Fe and Si and hence the relative importance of the two
supernova types at different group radii \citep{fino02}.  For the same
reasons, it is possible that the so-called Fe and Si biases
\citep{buot00a} may have affected some of these earlier results, at
least for groups with thermally complex core regions.
 
With the current generation of X-ray telescopes, the situation has
improved dramatically.  Recent {\em XMM} observations have shown that
temperature profiles of cool-core groups display a striking similarity
and typically peak at a radius $r \sim 0.05 r_{100}$ \citep{gast05},
which corresponds to $r \sim 0.1r_{500}$ for the mass profiles derived
for these groups.
Another recent study, also based on {\em XMM} observations, is the
work of \citet{fino07}, who investigated the distribution of
temperature, density, metal abundance, and derived quantities using
2-D spectral mapping of a sample of 14 groups.  This work showed that,
as for clusters, central metal excesses are present in groups with a
cool core, a result already hinted at in earlier {\em ASCA} studies
\citep{buot00a}.
 
While {\em Chandra} lacks the sensitivity of {\em XMM-Newton}, it has
a superior spatial resolution, enabling a more robust removal of point
sources, and also experiences a lower and more stable instrumental
background. {\em Chandra} data can therefore complement {\em XMM}
observations of groups, particularly in the group cores, but can also
serve as an important validity check on {\em XMM} results for the
X-ray faint outskirts of groups due to the lower background.  Here we
utilize this to derive radial profiles of temperature and the
abundance of Fe and Si from {\em Chandra} observations of a sample of
15 groups. Particular emphasis is given to combining the results
obtained for individual groups to explore statistical trends within
the sample.  One purpose of doing so is to obtain a clearer picture of
the content and behaviour of metals in group outskirts, where most of
the intragroup gas resides but where the low X-ray surface brightness
often prohibits robust constraints for individual systems. In
addition, this approach allows an investigation into the signatures of
feedback in group gas, including the origin of metals in SN and the
redistribution of gas by active galactic nuclei (AGN), in more detail
than is possible on a group-by-group basis.  Furthermore, many
numerical simulations of groups and clusters now include chemical
evolution of the ICM through the actions of galactic starburst winds
(e.g., \citealt{torn04, rome06, cora06}), AGN outflows \citep{moll07},
and galaxy--ICM interactions \citep{doma06}.  Therefore, it also seems
timely to provide a firm observational baseline against which to test
the input physics in such simulations, from the perspective of their
predictions for the chemical (and thermal) properties of the ICM in
the group regime.
 
The paper is structured as follows. In Section~\ref{sec,sample} we
describe the group sample and data analysis procedure, and
Section~\ref{sec,results} presents the main results for each group.
Section~\ref{sec,syst} discusses the impact of systematic errors on
our results. In Section~\ref{sec,combine} we stack the results for
individual groups, to investigate and quantify systematic trends
across the sample.  Section~\ref{sec,conclude} summarizes the results
and presents our conclusions. In a companion paper (Rasmussen \&
Ponman, in prep.; hereafter Paper~II), we discuss the implications of
these results for the history and nature of galactic feedback and
chemical enrichment within the sample.  $H_0=70$~km~s$^{-1}$~Mpc$^{-1}$
is assumed throughout, and all errors are given at the 68~per~cent
confidence level unless otherwise stated.

\section{Group sample and data analysis}\label{sec,sample} 
 
\subsection{Sample selection and data preparation} 
 
Our sample is predominantly based on the 25~GEMS groups from the
`G'-sample of \citet{osmo04a}. This subsample comprises those GEMS
groups that display group-scale X-ray emission (spatial extent of
$r\ga 60$~kpc), distinguished this way from systems that remain X-ray
undetected and those in which the diffuse emission can be associated
with the central galaxy rather than with the group as a whole. From
these 25 groups, we effectively imposed a flux limit on the sample by
only selecting systems with available {\em Chandra} archival data
containing more than 6,000 ACIS net counts from diffuse emission.
This was done in order to enable robust, spatially resolved
spectroscopy in a number of regions for each group.  In addition, only
groups more distant than 20~Mpc were included, to ensure that data
extended well outside the central group galaxy. As one of our aims is
to investigate signatures of galactic feedback in group gas, and since
we wish to study the radial distributions of gas properties, this
study requires systems which appear reasonably undisturbed by recent
merger activity and associated gas mixing. We therefore also excluded
NGC~5171 from the sample, an obviously unrelaxed system already
studied in detail by \citet*{osmo04b}.

This left 11 groups, to which we have added a further four groups with
data in the {\em Chandra} archive that conform to the same criteria.
The resulting sample includes two Hickson compact groups (HCG~42 and
62) and a fossil group (NGC~741) according to the definition of
\citet{jone03}, and spans a temperature range $kT=0.3$--2.1~keV, with
a median temperature of $T=1.06$~keV and median redshift of $z=0.014$.
It is worth emphasizing that the selected groups are all reasonably
X-ray bright and that the sample is by no means statistically
representative. In fact, the invoked net counts criterion alone
eliminated 13 of the 25 GEMS systems from which our sample was largely
drawn, resulting in a sample lower limit of $L_{\rm X}\approx 5\times
10^{41}$~erg~s$^{-1}$.  Three of our 15 groups are also included in
the {\em XMM} sample of \citet{fino06}, with a further 6 included in
the {\em XMM} sample of \citet{fino07}, enabling cross-comparison of the
results obtained from the two instruments.
  
For all our data sets, calibrated event lists were regenerated using
{\sc ciao} v3.3. For Very Faint mode observations, the standard
additional background screening was carried out.  Bad pixels were
screened out using the bad pixel map provided by the pipeline, and
remaining events were grade filtered, excluding {\em ASCA} grades 1,
5, and 7. Periods of high background were filtered using 3--$\sigma$
clipping of full--chip lightcurves, binned in time bins of length
259.8-s and extracted in off-source regions in the 2.5--7~keV band for
back-illuminated chips and 0.3--12~keV for front-illuminated chips.
Blank-sky background data from the calibration database were screened
and filtered as for source data, and reprojected to match the aspect
solution of the latter.
 
Point source searches were carried out with the {\sc ciao} task `{\sc
  wavdetect}' using a range of scales and detection thresholds, and
results were combined.
Source extents were quantified using the $4\sigma$ detection ellipses
from `{\sc wavdetect}', and these regions were masked out in all
subsequent analysis. Diffuse emission from individual group galaxies
other than the central early-type present in all groups was also
masked out.
 
To provide a rough idea about the spatial distribution of X-ray
emission in each system, images of all groups are presented in
Fig.~\ref{fig,mosaic}, showing the adaptively smoothed,
background-subtracted and exposure-corrected 0.7--2.5~keV emission on
the relevant central CCD (the ACIS-I array for NGC~6338 and NGC~7619,
ACIS-S2 for NGC~2300, and S3 for all others). These images were
generated following the procedure outlined in \citet*{rasm06}.
Although small-scale substructure is visible in some of the systems,
with the second brightest group galaxy also seen in X-rays in a few
cases, most of the groups are evidently reasonably dynamically
relaxed, supporting the use of one-dimensional radial profiles to
describe the spatial variation in hot gas properties of the systems.

\begin{figure*} 
\begin{center} 
 \mbox{\hspace{-1mm} 
 \includegraphics[width=178mm]{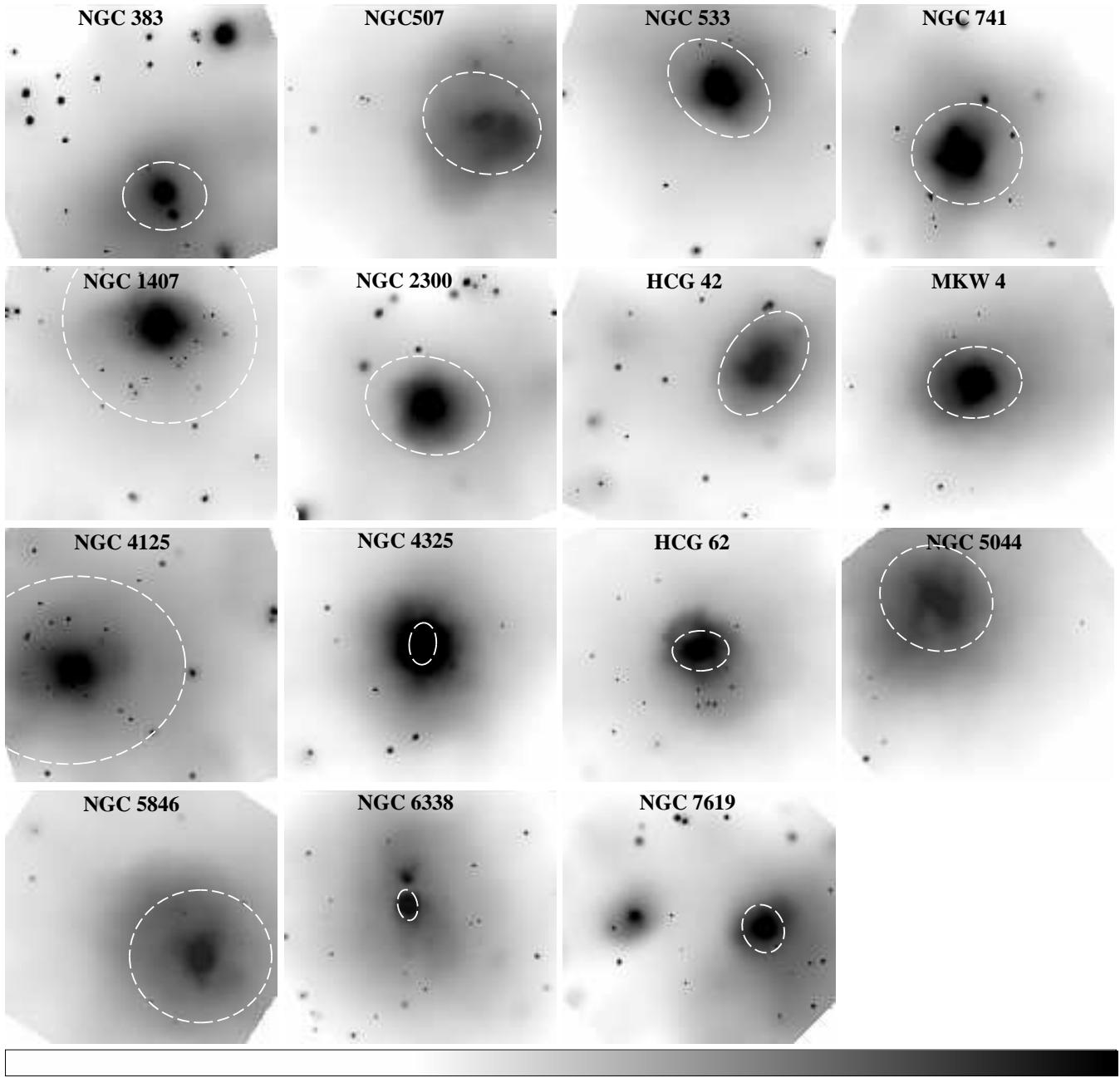}} 
\caption{Adaptively smoothed 0.7--2.5~keV images of the central region of
all groups.
  Images of NGC~6338 and NGC~7619 are 14.4~arcmin across, all others
  7.2~arcmin.  Dashed ellipses outline the $D_{25}$ ellipse of the
  brightest group galaxy.}
\label{fig,mosaic} 
\end{center} 
\end{figure*}

\subsection{Spectral fitting procedure}\label{sec,specfit}

All regions used for spectral extraction were centred on the peak of
the diffuse X-ray emission, which in most cases coincided to within
1--2~arcsec (and always within 5~arcsec) with the optical position of
the brightest group galaxy as listed in the Hyperleda database.  For
each CCD in each data set, the appropriate blank-sky data were scaled
to match off-source count rates in the 10--12~keV band and then used
to generate background spectra.  All spectra were fitted in {\sc
  xspec} v11.3 in the 0.7--5~keV band, with the low-energy cut-off
chosen such as to reduce contamination from residual Galactic soft
X-ray emission not accounted for by our blank-sky background
subtraction (see Section~\ref{sec,syst} for details). Indeed, no
systematic soft residuals between source and background spectra were
seen within this energy band for any of the groups.
 
To put all groups on an equal footing, all radii were normalized to 
$r_{500}$, using  
\begin{equation} 
  r_{500}(T) = 0.45 T^{1/2} \mbox{ $h_{70}^{-1}$~Mpc} , 
\label{eq,r500} 
\end{equation} 
which is an empirical relation based on data outside the cluster core
for a sample of 39 groups and clusters, including eight systems with
$T<1.5$~keV \citep*{fino01b}.  In order to obtain a
representative mean temperature $\langle T\rangle$ for use
with equation~(\ref{eq,r500}), as well as a corresponding mean
abundance $\langle Z\rangle$, and to create spectral weights for
subsequent fits in individual annuli, a source spectrum was first
extracted within the radial range 0.1--0.3~$r_{500}$.  This range was
chosen because it extends to the maximum radius covered by ACIS for
the more nearby systems, while in all cases being well outside the
optical extent of the central galaxy. The latter was quantified as
$D_{25}$, the ellipse outlining a $B$-band isophotal level of
25~mag~arcsec$^{-2}$.  The associated fit results should therefore
remain largely uncontaminated by X-ray emission from the interstellar
medium and any unresolved point sources in the central galaxy. As we
will see, this choice of radial range also excludes any cool core, in
accord with our usage of equation~(\ref{eq,r500}).
The fits to these spectra all adopted an absorbed thermal APEC model
in {\sc xspec}, assuming the abundance table from \citet{grev98}.
Using the derived value of $\langle T\rangle$ with
equation~(\ref{eq,r500}), $r_{500}$ was then re-evaluated iteratively,
until the resulting values of $\langle T\rangle$ varied by
less than 1~per~cent.  Throughout, we will refer to the values of
$\langle T\rangle$ and $\langle Z\rangle$ obtained in this way
as the mean temperature and abundance of each group.
 
Spectra were then extracted in concentric annuli containing at least
2000~net counts in the 0.7--1.5~keV band, and accumulated in bins of
at least 25~net counts.  Spectral response products were weighted by a
model of the `mean' spectrum extracted above.  Again, an absorbed APEC
model was first fitted to the data, yielding a mean temperature and
abundance for the relevant annulus. A second fit was then performed
using an absorbed VAPEC model, with all element abundances tied
together except for Fe and Si.
For spectra extracted inside the region covered by the central
early-type galaxy, a power-law of fixed index $\Gamma=1.72$ was added
to the spectral model, to account for any unresolved contribution from
low-mass X-ray binaries.
  
To investigate the presence of multiple thermal components and address
the importance of systematic errors associated with the Fe bias
\citep{buot00a}, we also fitted a 2-$T$ plasma model to all bins
inside the optical extent of the central galaxy, with both the Fe and
Si abundance tied for the two thermal components.
Where either the resulting Fe or Si abundance was significantly
different from the 1-$T$ result (at 90~per~cent confidence), we
adopted the best-fitting 2-$T$ values for both metals, confirming that
a 2-$T$ model provided a better fit in all cases.  As discussed in
more detail in Section~\ref{sec,syst}, this was relevant for only five
groups. We note that the adoption of the 2-$T$ results for the
abundance in these five cases does not affect $\langle Z \rangle$
significantly, as this quantity has been derived within
0.1--0.3~$r_{500}$, which is generally well outside the region for
which 2-$T$ models provide superior fits in the relevant groups.

Owing to the layout of the ACIS chips, we do not have full azimuthal
coverage of the groups at large angular distances from the group
centres. Our results at large radii could therefore be sensitive to
azimuthal variations in fitted parameters. In an attempt to
accommodate the systematic uncertainties associated with this, we used
azimuthal variations at smaller radii to gauge the variations further
out.  For each group, a number of spectra were extracted from an
annulus outside $D_{25}$ extending to the maximum radius of full
coverage. To mimic the situation at large radii, only data within a
narrow strip across this annulus were used, masking out the remaining
regions to achieve, if possible, 2000~net~counts within the strip, as
for the full-coverage annuli. This was repeated a number of times with
varying strip position angles. The resulting spectra were then each
fitted with a VAPEC model as described above, and the resulting
standard variations on relevant parameters ($T$, $Z_{\rm Fe}$, $Z_{\rm
  Si}$) adopted as the systematic uncertainties at large radii.  The
mean uncertainties resulting from this approach were 8, 34, and
58~per~cent on $T$, $Z_{\rm Fe}$, and $Z_{\rm Si}$, respectively.
These errors have been added in quadrature to the statistical errors
obtained for all annuli with less than 75~per~cent azimuthal coverage
(for most groups just one or two radial bins).

Table~\ref{tab,groups} records the position of the X-ray peak of each
group, group distance $D$ corrected for Virgocentric infall, the
useful ACIS exposure time $t_{\rm exp}$, $\langle T\rangle$
and $\langle Z \rangle$ from the `mean' spectrum, $r_{500}$ from
equation~(\ref{eq,r500}), and the Galactic value of the absorbing
column density $N_{\rm H}$ from \citet{dick90} as adopted in the fits.
Note that the statistical uncertainty on $r_{500}$ based on that of
$\langle T\rangle$ alone is typically on the order of just a
few per~cent. The actual uncertainties on $r_{500}$ are therefore
likely to be dominated by systematics associated with the use of
equation~(\ref{eq,r500}), but a quantification of this is beyond the
scope of this paper.

\begin{table*} 
 \centering 
 \begin{minipage}{129mm} 
   \caption{Properties of the group sample.} 
  \label{tab,groups} 
  \begin{tabular}{@{}lcccccccc@{}} \hline 
   \multicolumn{1}{l}{Group} & 
\multicolumn{1}{c}{RA} & 
\multicolumn{1}{c}{Dec} & 
\multicolumn{1}{c}{$D$} & 
\multicolumn{1}{c}{$t_{\rm exp}$} & 
\multicolumn{1}{c}{$\langle T \rangle$} & 
\multicolumn{1}{c}{$\langle Z \rangle$} & 
\multicolumn{1}{c}{$r_{500}$} & 
\multicolumn{1}{c}{$N_{\rm H}$} \\
  & (J2000)  & (J2000) & (Mpc) & (ks) & (keV) & (Z$_\odot$) & (kpc) &  
  ($10^{20}$ cm$^{-2}$) \\ \hline 
NGC~383  & 01 07 24.9 & $+32$ 24 45.8 & 73  & 42.5 & 1.65$^{+0.04}_{-0.06}$  
         & $0.39^{+0.07}_{-0.07}$ & 578 & 5.41 \\  
NGC~507  & 01 23 39.2 & $+33$ 15 19.6 & 69  & 14.3 & $1.30^{+0.03}_{-0.03}$  
         & $0.40^{+0.07}_{-0.06}$ & 513 & 5.23  \\   
NGC~533  & 01 25 31.4 & $+01$ 45 33.7 & 76  & 33.6 & $1.22^{+0.05}_{-0.05}$  
         & $0.28^{+0.07}_{-0.06}$ & 497 & 3.07 \\ 
NGC~741  & 01 56 21.0 & $+05$ 37 42.6 & 79  & 28.4 & $1.42^{+0.14}_{-0.12}$  
         & $0.16^{+0.08}_{-0.06}$ & 536 & 4.44 \\ 
NGC~1407 & 03 40 11.8 & $-18$ 34 48.9 & 26  & 37.7 & $1.01^{+0.07}_{-0.09}$  
         & $0.15^{+0.09}_{-0.06}$ & 452 & 5.42  \\ 
NGC~2300 & 07 32 19.2 & $+85$ 42 31.3 & 30  & 43.9 & $0.78^{+0.04}_{-0.03}$  
         & $0.17^{+0.07}_{-0.05}$ & 397 & 5.52 \\ 
HCG~42   & 10 00 14.2 & $-19$ 38 12.5 & 64  & 30.5 & $0.80^{+0.05}_{-0.05}$  
         & $0.25^{+0.31}_{-0.10}$ & 402 & 4.78 \\ 
MKW~4    & 12 04 26.9 & $+01$ 53 44.9 & 96  & 29.3 & $1.78^{+0.07}_{-0.09}$ 
         & $0.49^{+0.07}_{-0.07}$ & 600 & 1.88  \\ 
NGC~4125 & 12 08 05.7 & $+65$ 10 27.7 & 22  & 60.0 & $0.33^{+0.12}_{-0.05}$ 
         & $0.19^{+0.17}_{-0.07}$ & 259 & 1.82  \\ 
NGC~4325 & 12 23 06.7 & $+10$ 37 16.0 & 117 & 27.9 & $0.99^{+0.02}_{-0.02}$  
         & $0.39^{+0.08}_{-0.06}$ & 448 & 2.14 \\ 
HCG~62   & 12 53 05.8 & $-09$ 12 16.0 & 74  & 47.1 & $1.00^{+0.03}_{-0.03}$ 
         & $0.12^{+0.02}_{-0.03}$ & 450 & 3.06  \\ 
NGC~5044 & 13 15 24.0 & $-16$ 23 06.4 & 33  & 18.0 & $1.12^{+0.03}_{-0.03}$ 
         & $0.25^{+0.04}_{-0.03}$ & 476 & 4.94  \\ 
NGC~5846 & 15 06 29.4 & $+01$ 36 23.3 & 30  & 22.8 & $0.66^{+0.04}_{-0.03}$ 
         & $0.16^{+0.06}_{-0.05}$ & 366 & 4.24 \\ 
NGC~6338 & 17 15 22.9 & $+57$ 24 40.5 & 127 & 33.5 & $2.13^{+0.19}_{-0.07}$  
         & $0.25^{+0.08}_{-0.05}$ & 657 & 2.60 \\ 
NGC~7619 & 23 20 14.5 & $+08$ 12 24.9 & 52  & 25.6 & $1.06^{+0.07}_{-0.03}$ 
         & $0.23^{+0.05}_{-0.05}$ & 463 & 5.04 \\ 
\hline 
\end{tabular} 
\end{minipage} 
\end{table*}

\section{Results}\label{sec,results} 
 
In this section we present the results for the projected radial
profiles of temperature, Fe and Si abundance, and the ratio $Z_{\rm
  Si}/Z_{\rm Fe}$. As one of our aims is a statistical investigation
of radial trends across the sample, the profiles have not been
deprojected, since this would remove the statistical independence of
the individual radial bins. However, we discuss the impact of
deprojecting the data in Section~\ref{sec,syst}, showing that this
would not compromise our general conclusions.

In order to compare observed abundances and their ratios to those
expected from SN~Ia and SN~II enrichment, we adopted SN model yields
taken from the literature. For SN~Ia, the currently favoured
delayed-detonation WDD2 scenario is assumed \citep{iwam99}, while for
core-collapse supernovae (including types Ib and Ic), we have used the
recent yields of \citet{nomo06}. The latter were mass-averaged using a
Salpeter initial mass function (IMF) for the progenitors over the
range 10--50~M$_\odot$ (see \citealt*{fino03} and \citealt{depl07} for
a justification of this choice of IMF), so that
\begin{equation}
  M_i = \frac{ \int_{10 \mbox{{\tiny M}}_{\odot}}^{50 \mbox{{\tiny M}}_{\odot}}
    M_i(m) m^{-(1+x)} \mbox{ d}m}
  {\int_{10 \mbox{{\tiny M}}_{\odot}}^{50 \mbox{{\tiny M}}_{\odot}}m^{-(1+x)} 
    \mbox{ d}m}
\label{eq,SN}
\end{equation}
is the mass of the $i$-th element produced in a star of mass $m$, with
$x=1.35$. We have assumed a fractional metal mass of the SN~II
progenitors of $Z=0.004$ (i.e.\ $Z\approx 0.2$~Z$_\odot$), but the
exact choice has no strong bearing on the resulting Fe and Si yields
(see, e.g., \citealt{depl07}). The resulting average stellar yields
are $M_{\rm Fe}\approx 0.79$~M$_\odot$ and $M_{\rm Si}\approx
0.21$~M$_\odot$ for Fe and Si from SN~Ia, and $M_{\rm Fe}\approx
0.08$~M$_\odot$ and $M_{\rm Si}\approx 0.12$~M$_\odot$ for
core-collapse supernovae including SN~II. We note that the yield
uncertainties for SN~II are substantial, a factor $\sim 2$ for Fe and
$\sim 30$~per~cent for Si; the yields adopted here are bracketed by
the results of the different models considered by, for example,
\citet*{gibs97}, and so can be viewed as representative of typical
values in the literature. With the adopted abundance table, the
nominal abundance ratios expected for pure SN~Ia and SN~II enrichment
are $Z_{\rm Si}/Z_{\rm Fe}\approx 0.46$~Z$_{\rm Si,\odot}$/Z$_{\rm
  Fe,\odot}$ (SN~Ia) and 2.62~Z$_{\rm Si,\odot}$/Z$_{\rm Fe,\odot}$
(SN~II).  Note that this implies SN~II dominance (in absolute numbers)
for $Z_{\rm Si}/Z_{\rm Fe}\ga 0.65$, and that a solar abundance ratio
requires 3.4~SN~II per SN~Ia.

By adopting the SN~Ia yields of \citet{iwam99}, we restrict ourselves
to the `standard' picture of a reasonably homogeneous population of
SN~Ia for simplicity. We note, however, that recent developments
indicate that the situation could be more complex, with
evidence for a diversity in SN~Ia yields in cluster cores
\citep{fino02} and for two distinct populations of SN~Ia
\citep*{mann06}.

\subsection{Temperature profiles}\label{sec,temp}
Radial temperature profiles for all groups derived from
single-temperature models are presented in Fig.~\ref{fig,T}, with the
radii of all annuli normalized to $r_{500}$. It is worth noting that
emission is detected to fairly large radii in most of the groups, with
the midpoint of our outermost radial bin corresponding to roughly
75~per~cent of $r_{500}$.  This can be ascribed to the relatively low
and stable {\em Chandra} background.
 
\begin{figure*} 
 \includegraphics[width=176mm]{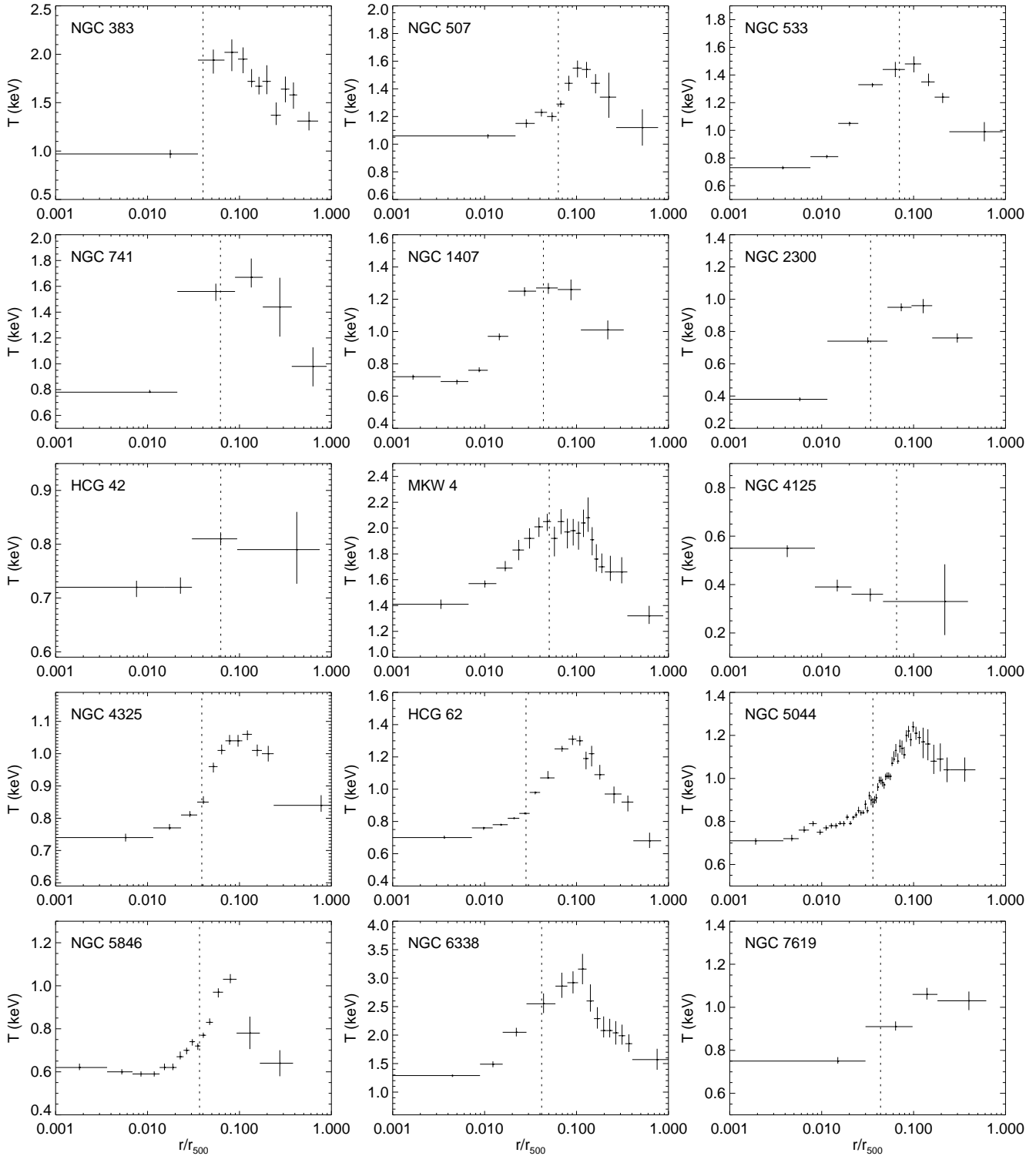} 
 \caption{Projected radial temperature profiles of all groups. Vertical dotted
   lines mark the semi-major axis of the $D_{25}$ ellipse of the
   central group galaxy.}
\label{fig,T} 
\end{figure*} 

In general, the temperature profiles appear remarkably similar, with
all but one system (NGC~4125) showing evidence for cooler gas in the
central regions. The temperature profiles typically peak around $r
\sim 0.1r_{500}$ and decline at large radii. There is also an
indication that the profiles flatten in the central regions, in
general agremeent with the group results of \citet{fino07}. Compared
to analogous profiles for relaxed clusters, this contrasts with the
projected {\em Chandra} temperature profiles of \citet{vikh05}, the
deprojected ones of \citet{sand06}, and the deprojected {\em XMM}
profiles of \citet{piff05}, for which there is no clear evidence for a
flattening at small radii. Defining the gas entropy as
$S=Tn_e^{-2/3}$, the deprojected cluster entropy profiles of
\citet{dona06} {\em do} show a central flattening, but this was
attributed to a central core in the density distribution rather than a
flattening of the temperature profile.
 
The typical location of the temperature peaks in our sample is in
excellent agreement with the {\em XMM} results of \citet{gast05}, who
also found temperature peaks at $r \sim 0.1 r_{500}$ for their
cool-core groups and poor clusters. Their sample has four groups in
common with ours (NGC~533, MKW~4, NGC~4325, and NGC~5044), and has
$r_{500}$ evaluated directly from the mass profiles derived for each
group.  The good agreement with our results therefore also suggests
that our estimates of $r_{500}$ are generally not seriously biased.
As also noted by \citet{gast05} for their sample, the temperatures
peak at larger overdensity than in clusters, where the profiles turn
over at $r \approx 0.15r_{500} \approx 0.1r_{200}$
\citep{piff05,vikh05,zhan06,sand06}. It is clear, however, even when
normalized to $r_{500}$, that the exact radius $r(T_{\rm max})$ at
which the peak occurs (in other words, the radial extent of the cool
core) varies slightly from group to group. Naturally, this result will
to some extent depend on the spatial resolution resulting from our
binning criterion and thus the surface brightness of the group
emission, but we note that the adopted spatial resolution does not
depend systematically on, for example, group temperature. We discuss
the location of the temperature peak in more detail in
Section~\ref{sec,temp2}.

\subsection{Abundance profiles}

The abundance profiles of Fe and Si are plotted in
Fig.~\ref{fig,abund}. In all systems except NGC~2300, and possibly the
non--cool core group NGC~4125, there is clear evidence for a central
excess of iron, as also seen for cool-core clusters. This central
excess often extends well beyond the optical extent of the central
galaxy, and in some cases shows an off-centre peak inside $D_{25}$,
with a decline in abundance in the very centre.  NGC~533 is perhaps
the clearest example of this, but the majority of groups with a
well-resolved core (roughly 8/11) shows some evidence of this feature.
This Fe peak is typically accompanied by a similar feature in the Si
distribution, most clearly in HCG~42, MKW~4, and HCG~62, with a hint
of similar behaviour in a few more groups.  Potentially, the peak
could be ascribed to Fe/Si bias inside the central galaxy, lowering
the derived abundance in the very core, but as discussed below, we
find little evidence that this effect is important for our analysis in
general. In addition, such a peak is commonly seen also in
well-resolved cluster cores where the Fe bias is generally expected to
be less important, e.g., Centaurus \citep{sand02}, Perseus
\citep*{sand05}, A1795 \citep{etto02}, and A2199 \citep{john02}.
Alternative explanations for the peak include AGN redistribution of
enriched gas (e.g., \citealt{math03,rebu06}), resonant scattering of
emission line photons in group cores \citep*{gilf87}, or gravitational
sedimentation of helium nuclei in the core \citep{etto06}. We will
discuss these possibilities in detail in Paper~II.

At large radii, a common feature of the Fe distributions is a drop in
$Z_{\rm Fe}$ to a value of $\approx 0.1$~Z$_\odot$. In some groups, a
rise in $Z_{\rm Si}$ is also seen, largely independent of the
behaviour of the Fe profile. We note that there is no indication that
derived profiles for the two Hickson groups or the fossil group
NGC~741 differ significantly from those of the rest of the sample in
this context.

\begin{figure*} 
\mbox{\hspace{-7mm} 
 \includegraphics[width=93mm]{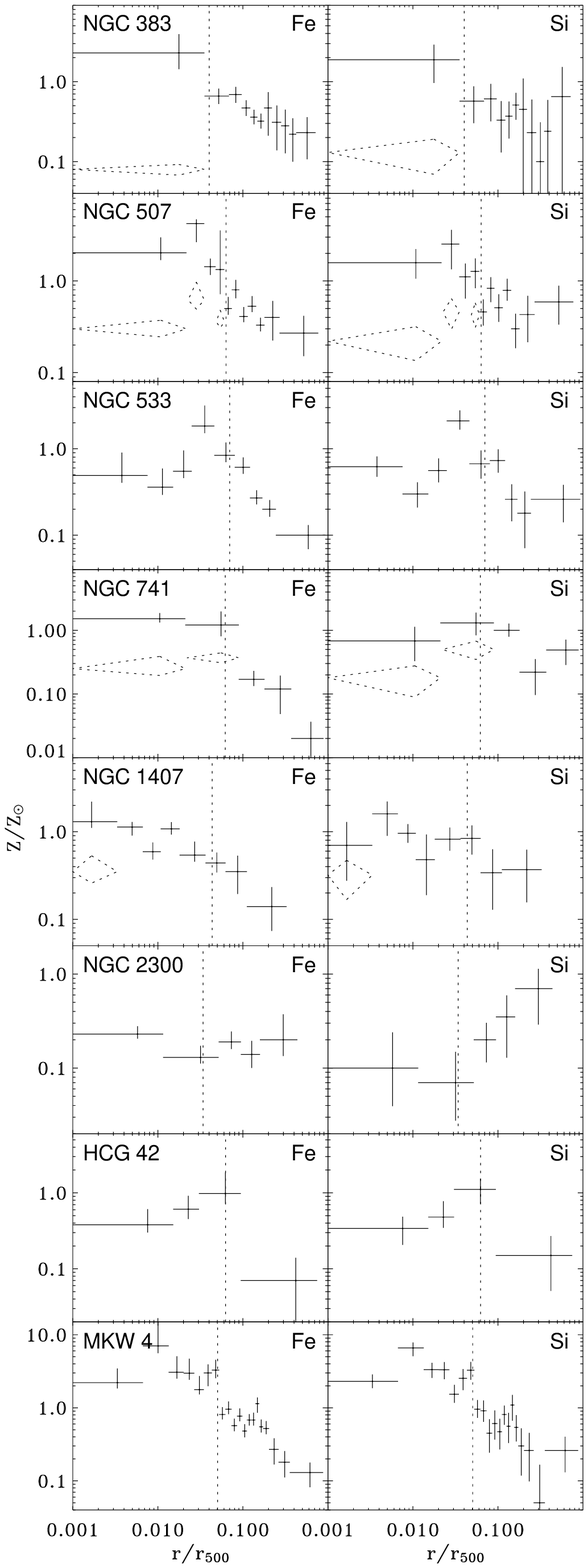}
 \includegraphics[width=93mm]{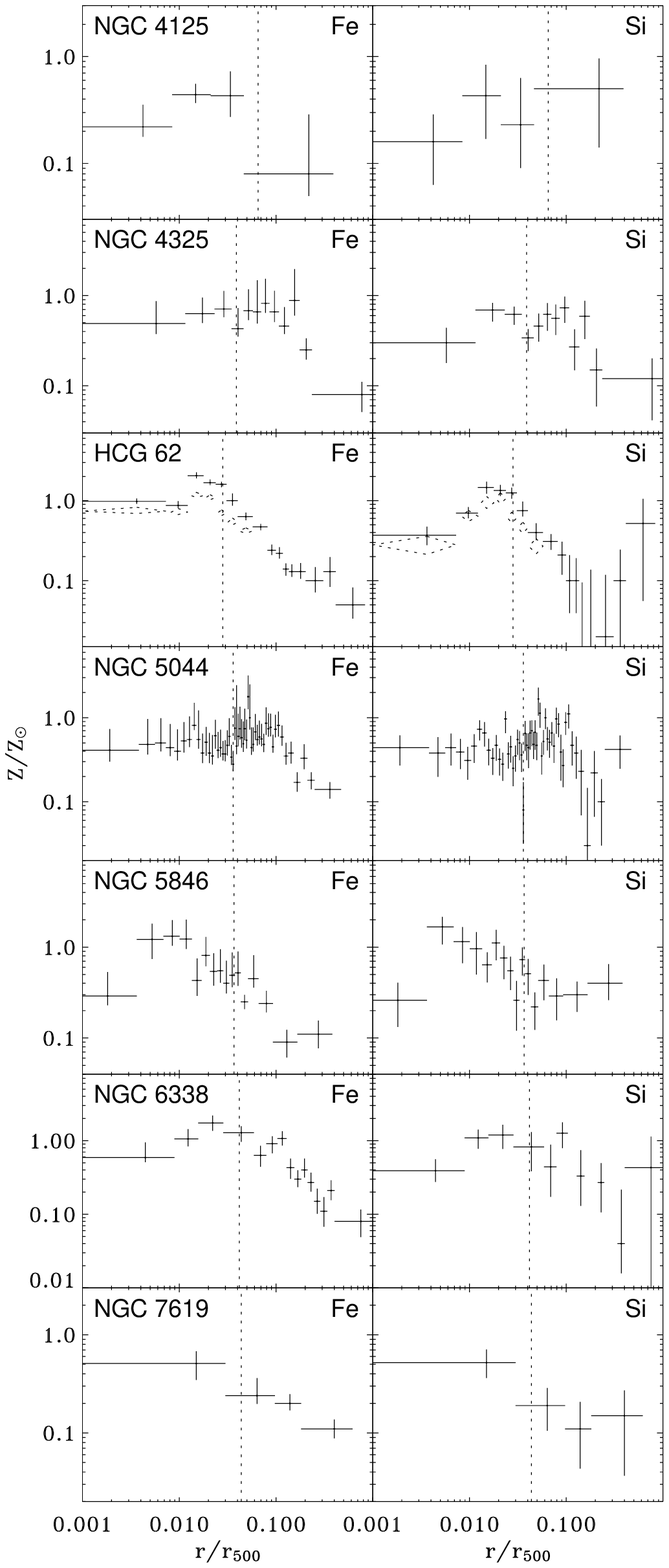}} 
\caption{As Fig.~\ref{fig,T}, but for the Fe and Si profiles. Where
  relevant (see Section~\ref{sec,specfit}), the result of fitting
  two-temperature models have been plotted, with the corresponding
  1-$T$ results shown as dotted diamonds.}
\label{fig,abund} 
\end{figure*}

The silicon-to-iron ratio $Z_{\rm Si}/Z_{\rm Fe}$ is plotted for all
groups in Fig.~\ref{fig,ratio}. In many cases the ratio in the central
bin is seen to be consistent with the Solar ratio, with a contribution
from SN~II formally required for about half the groups. For the other
half, there is clear evidence for subsolar abundance ratios,
consistent with enrichment by SN~Ia alone.  We do not believe there is
any fundamental difference between the two subsamples if split
according to this, as $Z_{\rm Si}/Z_{\rm Fe}$ in the central bin
clearly spans a continuous range of values rather than showing a
bimodal distribution. Given the variation in spatial resolution among
the groups and the uncertainties on $Z_{\rm Si}/Z_{\rm Fe}$, we
caution further against such an interpretation at this stage. For
example, we find no evidence that the two subsamples cluster
differently in the parameter space spanned by $\langle T\rangle$,
$\langle Z\rangle$, and $K$-band luminosity of the central galaxy.
Nevertheless, we will revisit this issue in Paper~II on the basis of
integrated Fe and Si masses inside $D_{25}$.

As discussed by \citet{fino02}, there is observational evidence for
diversity in SN~Ia yields in the cool cores of more massive clusters,
with solar Si/Fe ratios arising entirely from SN~Ia.  While our
results allow for this possibility, we cannot easily confirm that this
result extends down into the group regime, as it is based on the
abundances of additional elements such as S, Ca, and Ar.  At the lower
temperatures prevalent in groups, the abundances of these elements,
with prominent emission lines in the range $E \approx 2.5-5$~keV, are
not in general robustly constrained.  In the outer regions, the
interpretation of our data remains unaffected by this; here a
contribution from SN~II is universally required within our sample, and
the abundance ratios, given the typical uncertainties in SN~II yields
of a factor $\sim 2$ (e.g.\ \citealt{gibs97}), are consistent with
pure SN~II enrichment in all cases.
   
\begin{figure*} 
\mbox{\hspace{-0mm} 
 \includegraphics[width=176mm]{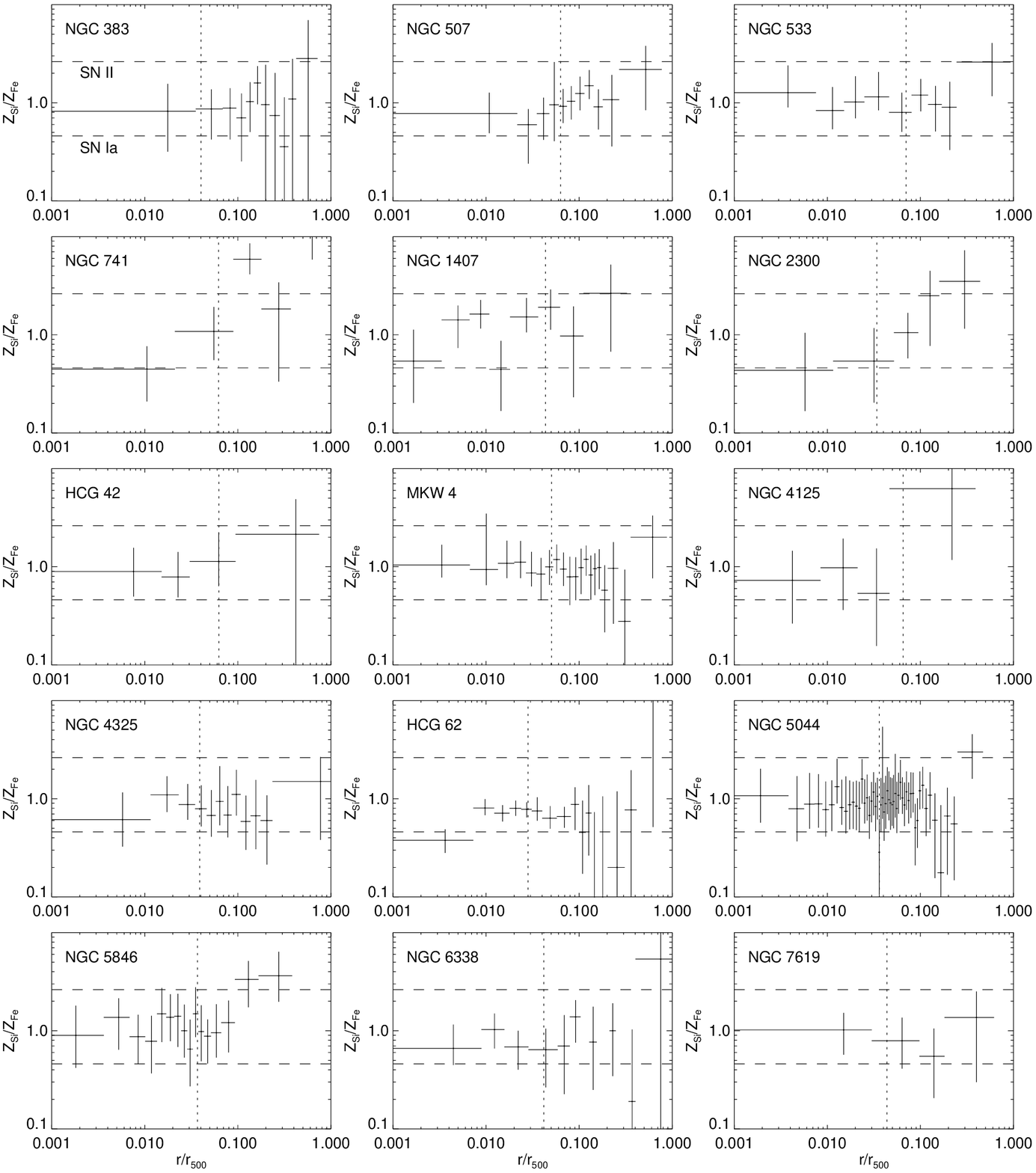}} 
\caption{As Fig.~\ref{fig,T}, but for the ratio of $Z_{\rm Si}/Z_{\rm
    Fe}$ (in units of Z$_{{\rm Si},\odot}$/Z$_{{\rm Fe},\odot}$). The
  lower and upper dashed lines mark the ratios expected from pure
  SN~Ia and SN~II enrichment, respectively.}
\label{fig,ratio} 
\end{figure*}

\subsection{Notes on individual systems}\label{sec,notes} 
 
Many of the groups exhibit properties which warrant some additional
comments. Below, we briefly discuss each group in the sample in turn,
with emphasis on the derived abundances.  Where relevant, we have also
attempted a comparison of the derived Fe profile to existing
measurements, in particular the 2-$D$ {\em XMM} results of
\citet{fino06,fino07} (F06 and F07 in the following), while correcting
for differences in adopted abundance tables.
It should be noted, however, that due to our incomplete azimuthal
coverage at large radii, direct comparison is not always
straightforward for sources with some degree of asymmetry in the Fe
distribution.\\

\noindent{{\bf NGC~383}}: 
The central galaxy hosts the bright radio source 3C~31, representing
one of the few examples of a central group galaxy showing a clearly
identified large-scale radio jet with a well-constrained jet power
\citep{lain02}. Despite this, the $T$-- and $Z$-profiles of this group
do not exhibit any clear distinguishing features compared to the rest
of our sample. For the temperature profile, this result is in accord
with the group results of \citet{jeth07}, which show that the slope of
$T(r)$ inside the cool core does not show any strong dependence on the
radio power of the central radio source.\\

\noindent{{\bf NGC~507}}: An example of a group in which radio lobes
are present around the central galaxy \citep{kraf04} and in which the
highest-metallicity gas is displaced from the group core, perhaps as a
consequence of radio outbursts lifting enriched gas out of the core.
If so, one might expect to see some degree of asymmetry in the Fe
distribution in the core. The 2-$D$ Fe map of F07 does reveal some
asymmetry inside $0.05r_{500}$, showing a high-metallicity blob
south-west of the group core, but it is not immediately obvious that
this is necessarily associated with an AGN outflow. To achieve a
satisfactory fit in the outermost radial bin in our data, it was
necessary to exclude the second brightest group galaxy, NGC~499, out
to 6~arcmin ($\sim 120$~kpc). Comparison to the single-$T$ Fe profile
of F07 shows good overall agreement, including the off-centre Fe peak
at $\sim 0.03r_{500}$ and the decline to $\sim 0.3$~Z$_\odot$ at
$0.5r_{500}$.\\

\noindent{{\bf NGC~533}}: This group presents one of the clearest
examples of an off-centre peak in the Fe distribution within our
sample.  As for NGC~507, which also hosts such an Fe peak, 1.4-GHz
NVSS data \citep{cond98} reveal a central radio source.  Inside
$0.02r_{500}$ our Fe results are consistent with the F07 value of
$\sim 0.7$~Z$_\odot$, as is our detection of the supersolar Fe peak at
$0.04r_{500}$. The clear decline to 0.1~Z$_\odot$ at large $r$ is not
obvious from the F07 results, however, which show considerable scatter
outside $0.1r_{500}$. We speculate that this relates to our assumption
of circular symmetry and our incomplete azimuthal coverage at large
radii.  The 2-$D$ Fe map of F07 suggests an elongated abundance
structure on large scales, with the ACIS-S array in our {\em Chandra}
data running very nearly perpendicular to the major axis of this
structure. This implies that our result includes low-metallicity gas
at the largest radii, beyond those included in the F07 analysis, and
that we are inevitably missing
some of the higher-metallicity gas seen by F07.\\

\noindent{{\bf NGC~741}}: With a group X-ray luminosity of $3\times
10^{42}$~erg~s$^{-1}$ \citep{osmo04a}, and a $B$-band magnitude
difference between the first and second brightest galaxy (the spiral
UGC~1435) within $r_{500}$ of $\Delta m_{12}=2.6$, this system meets
the criteria for fossil groups introduced by \citet{jone03}.  Its
temperature and metallicity profiles appear similar to those of most
other systems studied here, though the Fe abundance at large radii is
clearly lower than average.  It is also a group with a powerful
central radio source and large-scale radio emission (e.g.,
\citealt{jeth07}), although the latter is probably a tailed radio
source associated with the nearby group member NGC~742 rather than
with the central elliptical itself (Jetha et al., in prep.).\\

\noindent{{\bf NGC~1407}}: This is probably a dynamically evolved
group, with overall optical properties not too dissimilar to those of
the fossil system NGC~741 \citep*{tren06}. Its Fe and Si profiles
appear fairly typical for the sample.\\

\noindent{{\bf NGC~2300}}:
This is the only group within our sample with a spatially resolved Fe
profile outside $D_{25}$ that does not show clear evidence for a
radial decline. The Si profile is also somewhat atypical of the
sample, showing a clear rise with radius outside the core. The
relatively low and near--constant Fe abundance, as well as the radial
rise in Si, is in very good agreement with earlier {\em ROSAT} and
{\em ASCA} results \citep{davi96,fino02}, although the {\em XMM} study
of F06 seems to find slightly higher Fe abundances inside $0.3r_{500}$
than those seen here.
With the central early-type galaxy (NGC~2300 itself) having a stellar
mass $\sim 10$~times that of the second largest galaxy, NGC~2276
\citep{rasm06}, the central galaxy could potentially have been almost
solely responsible for the chemical enrichment of the hot gas in this
system. This is also an example of a group in which ram-pressure
stripping of (presumably enriched) galactic gas is taking place
\citep{rasm06}, illustrating that an SN--driven outflow is not the
only mechanism for enriching the hot gas in the group. We note that
there is also evidence of a recent merger in the central early-type
\citep{forb92}, with further evidence for this seen in the ICM
pressure map of \citet{fino06}.

Given that other features of this group seem typical of our sample,
such as the presence of a cool core and a fairly undisturbed X-ray
morphology, the question remains how to interpret the uniformly low Fe
abundance and the rise in $Z_{\rm Si}$ with radius. We note that
$Z_{\rm Fe}$ at large radii is not atypical of the sample, nor is the
radial profile of Si/Fe. This suggests a suppression of both Fe and Si
enrichment in the group core. In light of the evidence of a recent
merger in the central early-type, one possible explanation is that a
cool, low-$Z$ system has dropped into the group core, thus diluting
any Fe excess but preserving the cool core.

Another possibility, potentially induced by the very low central
temperature, is that high-$Z$ gas has cooled out in the group core.
Based on the measured X-ray flux, the fitted spectral parameters, and
the gas density profile of \citet{davi96}, we estimate a central
cooling time of only $\sim 1\times 10^8$~yr. This would render such
cool-out a viable scenario, in particular in light of the absence of
any strong radio source in the central early-type (1.4~GHz radio power
$P\approx 5\times 10^{20}$~W~Hz$^{-1}$ based on the NVSS flux) which
could indicate ongoing AGN heating activity.  In order to remove any
Fe excess and explain the `suppressed' Fe level in the core, a total
Fe mass of $2.5\times 10^5$~M$_\odot$ must have cooled out inside the
typical radial extent of the Fe excess of $r\approx 0.05r_{500}$,
assuming an initial Fe abundance of $\sim 0.7$~Z$_\odot$ (see
Section~\ref{sec,combine} for this choice of parameters).  This is
equivalent to a total gas mass of $2.0\times 10^8$~M$_\odot$, which
does not seem an unreasonably large amount.  In addition, cooling
would also have raised the central entropy through the removal of
low-entropy gas from the X-ray phase. Among the nine of our groups
also included in the F06 and F07 samples, NGC~2300 does show the
highest central entropy, lying $1.6\sigma$ above the sample mean
derived inside $0.1r_{500}$ (F06). One concern for the cooling
scenario, however, is that the cooled gas is likely to participate in
star formation \citep{edwa07}, but there is no indication that the
central early-type is unusually blue, displaying a $B$--$K$ colour of
4.2. Based on the X-ray data alone, we conclude that the cooling
scenario provides a possible explanation for the lack of an Fe excess
in this group, but further tests of this would require constraints on
the amount of cold gas and young stars in the central early-type galaxy.\\

\noindent{{\bf HCG~42}}: The observed range of $Z_{\rm Fe} =
0.5-1.0$~Z$_\odot$ inside $0.1r_{500}$ is consistent with the F07
results, as is the decline at large radii. However, finding a low
value of $\sim 0.1$~Z$_\odot$ outside the cool core, we do not confirm
the value of $0.2$~Z$_\odot$ seen in the outermost bin of F07,
certainly in part because we do not have the statistics required to
resolve this narrow region around $0.5r_{500}$.\\

\noindent{{\bf MKW~4}}: The presence of an off-centre Fe peak in this
group was already claimed by \citet*{fuka04}, who furthermore noted
that no radio activity or merger evidence has been reported for the
system. Indeed, NVSS data show no evidence of a central radio source,
nor of any larger-scale radio emission surrounding the central galaxy,
NGC~4073 \citep{jeth07}.  This is in clear contrast to NGC~507 and
NGC~533, where this Fe peak is accompanied by the presence of a
central radio source. We confirm the high central Fe abundance of
$\sim 2$~Z$_\odot$ seen by F07, along with the sharp decline, starting
at $0.05r_{500}$, towards a value of 0.1~Z$_\odot$ at large radii.\\

\noindent{{\bf NGC~4125}}: This group is the lowest-temperature system
in the sample. It shows no cool core, despite appearing reasonably
relaxed in the X-ray (Fig.~\ref{fig,mosaic}).
Subdividing the innermost radial bin and fitting spectra for each
sub-bin, each only $\approx 500$~pc wide, still does not reveal any
temperature drop. As there appears to be a point-like X-ray source at
the centre, detected above the local background at $\sim 2\sigma$
significance, one possibility is that the central temperature rise
could be due to recent AGN heating. However, although a radio source
is present in NVSS data inside $D_{25}$, this source is displaced from
the optical and X-ray centre by 0.8~arcmin (4~kpc). To investigate the
nature of the central X-ray source, its spectrum was fitted with a
power-law absorbed by the Galactic value of $N_{\rm H}$. This provided
an unacceptable fit, even when allowing for intrinsic absorption
beyond the Galactic value (red.\ $\chi^2 > 1.7$ for 9 degrees of
freedom).  In contrast, an APEC plus power-law model fits well (red.\
$\chi^2 \approx 0.6$), yielding $T=0.56\pm 0.12$ and
$Z=0.39^{+0.34}_{-0.18}$ for the thermal plasma, and $\Gamma \approx
1.7$ for the power-law, with a total, unabsorbed luminosity of
$1\times 10^{39}$~erg~s$^{-1}$.  The plasma parameters are in good
agreement with those found on larger scales in the group core, and the
power-law with a contribution from low-mass X-ray binaries in the
central elliptical. The {\em Chandra} detection of a central point
source might therefore simply be attributed to a strongly peaked
surface brightness profile, in line with the {\em ROSAT} finding of
\cite{mulc03} for this group. We therefore conclude that there is no
strong X-ray evidence for ongoing AGN heating having raised the
central ICM temperature. Further support for this interpretation is
provided by the low X-ray luminosity derived for the very central
regions, which is well below typical AGN luminosities.

An alternative explanation for the central temperature rise could be
sought in the fact that the central elliptical is `dynamically young'
\citep{fabb95}, so a merger event within the past few Gyr could have
disturbed the central regions and mixed in hotter gas from larger
radii. The derived central cooling time is very short, however, $\sim
1\times 10^8$~yr as for NGC~2300, but increasing to more than 3~Gyr at
$D_{25}$ as a consequence of the sharply peaked surface brightness,
and hence density, profile \citep{mulc03}.  As this is a factor of
three above the sound crossing time within that region, it may suggest
that the gas density distribution has largely recovered from such a
merger disturbance, but significant central cooling has yet to
commence.

Another possibility is that the situation in NGC\,4125 is somewhat
similar to that seen for the fossil group NGC\,6482, one of the
coolest known fossils, with a temperature of only $\approx 0.5$~keV
outside the central 10~kpc \citep*{khos04}. Despite looking relaxed in
the X-ray and displaying a central cooling time of $\la 10^8$~yr, this
system also shows no cool core. A possible explanation is that any gas
taking part in a steady-state cooling flow in this group is actually
{\em heated} by gravitational $P$d$V$ work as it moves inwards, owing
to the very cuspy total mass profile of the group. \citet{khos04} find
that, in such a scenario, the observed temperature profile can be
explained by a cooling-flow mass accretion rate of just
2~M$_\odot$~yr$^{-1}$. The sharply peaked density profile seen for
NGC~4125 is consistent with the idea that a similar mechanism could be
at work in this group.\\

\noindent{{\bf NGC~4325}}: This is a group in which the extent of the
metal excess in the core substantially exceeds the optical extent of
the central galaxy. No central radio source is seen in NVSS data
\citep{jeth07}, but there is evidence of a previous weak AGN outburst,
with an indication that another outburst may be about to occur
\citep*{russ07}. The observed near-constancy of $Z_{\rm Fe}\approx
0.7-0.8$~Z$_\odot$ inside $0.1r_{500}$ agrees well with the F07
results, as does the very steep decline to a value $Z_{\rm
  Fe}<0.1$~Z$_\odot$ at $0.8r_{500}$. \\

\noindent{{\bf HCG~62}}: This group has been extensively studied with
{\em ROSAT}, {\em ASCA}, {\em Chandra}, and {\em XMM}
\citep{fino99,buot00b,mori06}.  It harbours one of the clearest
examples of X-ray cavities in groups \citep{vrti00,mori06}, indicating
past radio activity. The older {\sc ciao} task `{\sc mkrmf}' was used
to create spectral responses for this group, due to a lack of relevant
calibration data for the newer task `{\sc mkacisrmf}' otherwise used
in this study. Despite this and the complex X-ray morphology in the
core of this system, the combined {\em Chandra} and {\em XMM} study of
\citet{mori06} finds a temperature profile very similar to the one
derived here, and the roughly solar Fe and Si abundances seen in the
core agrees well with the results of \citet{buot00b} and
\citet{mori06}.  The sharp decline in the Fe profile, beginning at
$\sim 0.03r_{500}$ and reaching a value of $\sim
0.05$~Z$_\odot$ at $0.6r_{500}$, is also in good agreement with F07.\\

\noindent{{\bf NGC~5044}}: This system has been studied extensively
with both {\em XMM} and {\em Chandra} \citep*{buot03a,buot03b,buot04}.
At $r\approx 0.05r_{500}$, our results are consistent with the solar
Fe abundance found by F06.  In our outermost radial bin, we find
$Z_{\rm Fe} = 0.14\pm 0.04$~Z$_\odot$, confirming the low Fe abundance
of 0.1--0.15~Z$_\odot$ derived at large radii by 
\citet{fino99} and \citet{buot04}.\\

\noindent{{\bf NGC~5846}}: The Fe map of F06 suggests a highly
symmetric abundance structure, and supports our detection of an
off-centre Fe peak and a decline to 0.1~Z$_\odot$ at large radii.\\

\noindent{{\bf NGC~6338}}: Reaching a peak temperature of $T\approx
3$~keV, this is the hottest system in our sample. In
Fig.~\ref{fig,mosaic}, extended emission is also seen from the second
brightest galaxy. This has been masked out to a radius of 1~arcmin in
all spectral fits. As for NGC~4325, the central metal excess in this
group extends well beyond the central galaxy, but is here also
accompanied by a central radio source. \\

\noindent{{\bf NGC~7619}}: NGC~7619 itself is one of a central pair of
ellipticals of similar optical brightness in the group.  Extended
X-ray emission is also seen in Fig.~\ref{fig,mosaic} around the other
pair member, NGC~7626; this was masked out to a radius of 2~arcmin in
all spectral fits.

\section{Systematic errors}\label{sec,syst} 

Before proceeding to discuss the results in more detail, we address a
number of analysis issues which could be affecting our results. These
include the presence of Fe and Si biases in the group core, and the
reliability of our background estimates.  An immediate, although
indirect, indication that such issues are not having a major impact on
our results, is the fact, discussed below, that we detect the same
overall radial trend in $T$ and $Z$ for all systems, despite
considerable variations in both source and background flux across the
sample. For the temperature profiles, the agreement with the {\em
  XMM}~result of \citet{gast05} for the mean position of the
temperature peak is encouraging. Moreover, there is good overall
agreement with derived Fe profiles for the six groups overlapping with
the {\em XMM} sample of \citet{fino07}, even at large radii where
systematic uncertainties related to background subtraction could be
important.

\subsection{Geometric effects}

One issue which could call the robustness of our results into question
is the fact that our profiles are projected ones, so all bins contain
some contribution from emission further out. To test the importance of
this effect, we deprojected the profiles for a few groups using the
{\sc projct} model in {\sc xspec}. For simplicity, a 1-$T$ APEC model
was fitted to the data. The result for NGC~4325 is shown in
Fig.~\ref{fig,projct}. This group is typical of our sample in terms of
the number of radial bins and the value of $\langle T
\rangle$; it also shows a near-solar Si/Fe ratio throughout, so the
measurements should not be biased strongly by fixing the relative
abundances. The figure demonstrates reasonable overall agreement
between projected and deprojected results. As can be seen, projection
tends to slightly reduce the temperature gradient in the core, as
would be expected. While the deprojected abundance profile recovers
the increased Fe level in the region inside $\sim 0.1r_{500}$,
significant fluctuations between adjacent bins are evident in this
region in disagreement with the projected profile.  These are most
likely artefacts related to issues of fit stability, a problem
commonly arising for spectral deprojections involving a large number
of bins (e.g., \citealt{sand06}). Results for less well-resolved
systems support this suspicion and confirm the lack of systematic
variations between projected and deprojected results. We therefore
conclude that our results are reasonably reliable representations of
the three-dimensional profiles.

\begin{figure} 
\begin{center} 
\hspace{-5mm}
 \includegraphics[width=80mm]{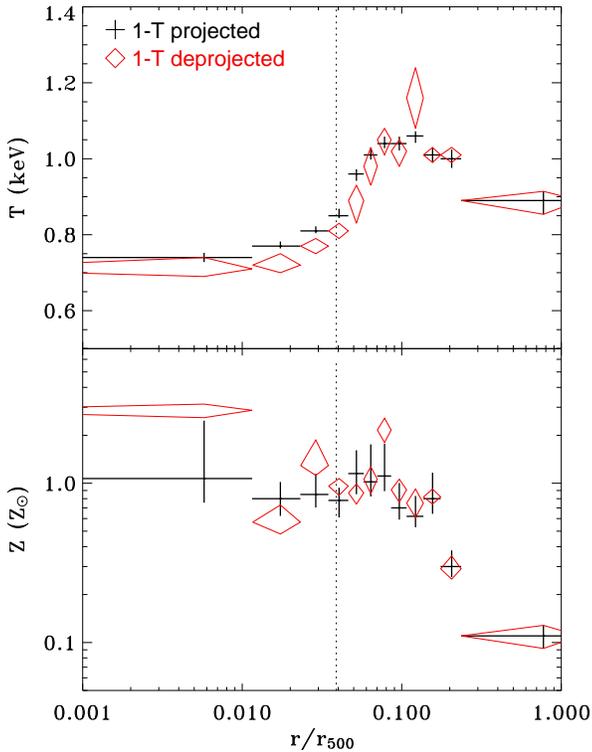} 
 \caption{Projected and deprojected temperature and abundance profiles 
   for NGC~4325, resulting from fitting a single-temperature model to 
   all bins.} 
\label{fig,projct} 
\end{center} 
\end{figure} 
 
Another issue is the fact that we have incomplete azimuthal coverage
at large radii. This could potentially play a role for individual
groups in which a simple radial description of the abundance profile
breaks down in the outskirts. As mentioned, however, we have attempted
to select systems with a reasonably regular X-ray morphology, and we
have also tried to accommodate any remaining uncertainties related to
this in our error analysis. The good overall agreement with {\em XMM}
results at large radii indicates that this effect is not important, at
least for the groups where a comparison is possible. In addition, any
bias would be strongly suppressed when averaging the results across
the group sample, which is the approach taken in the next Section. We
therefore believe our conclusions to be robust against this effect.

\subsection{Observational bias in group cores}

It is well known that high-quality observations of groups in many
cases reveal disturbed central structures, showing that group cores
can be thermally complex (e.g.\ \citealt{kim04,osul05}) and require
multi-component thermal models to properly fit their spectra.  To
investigate this, we plotted the distribution of reduced $\chi^2$
values for all our fits, but found no clear trend with radius and no
indication that multi-temperature models are generally {\em required}
to describe the central regions.

Nevertheless, particularly for systems such as NGC~383, where the
central cool regions are not well resolved because of our binning
criterion, the presence of multi-phase gas could lead to artificially
low abundance estimates in the innermost bin \citep{buot00a}, and
metallicities derived from 1-$T$ fits should here be regarded with
some caution.  We tested for this effect by first radially subdividing
the central radial bin and fitting spectra extracted within each of
these two sub-bins.  Eight of the 15 groups show evidence for $T$
variations between those two sub-bins.  For all systems, the abundance
values within the two sub-bins were consistent with the full-bin
value, and only one group displays evidence for $Z$ variation between
the two sub-bins (NGC~383, with the innermost sub-bin showing a lower
value).  However, this does not in itself rule out the presence of
multi-phase gas within each of these two sub-bins. A further test
therefore involved fitting two-temperature models to all bins within
$D_{25}$, as already described in Section~\ref{sec,specfit}. Only for
five groups do we find evidence that 2-$T$ models yield significantly
higher abundances {\em and} provide superior fits, in some of the
innermost radial bins (Fig.~\ref{fig,abund}). For these, the 2-$T$
results for the abundances have been adopted, as plotted in
Fig.~\ref{fig,abund}. An example is NGC~507, for which only the 2-$T$
fit recovers the high Fe abundance of 2--3~Z$_\odot$ reported inside
$D_{25}$ by \citet{kim04} on the basis of {\em XMM} data.
 
While 2-$T$ models certainly provide better fits in some cases, 
it is perhaps slightly surprising that they only yield significantly
different abundances in a few of these, given the presence of central
temperature variations in most of our groups. A possible explanation
is that we are typically sampling the core regions with rather finely
spaced annuli. The limited radial extent of each annulus may imply
that mixing of temperature components may not be important within each
bin. For example, for the very core of MKW~4, \citet{fuka04} also find
from the same {\em Chandra} data that 2--$T$ models do not produce
significantly improved fits or higher $Z_{\rm Fe}$, in agreement with
our results. A different conclusion was reached by \citet{osul03}
based on {\em XMM} data, which \citet{fuka04} attribute to the much
larger spectral extraction region used by the latter authors. Support
for this interpretation is provided by the work of \citet{hump06},
which shows that {\em Chandra} spectra of elliptical galaxies can
generally be described by 1-$T$ models, and that the multi-$T$ fits
required to describe earlier {\em ASCA} spectra of the same objects
were an artefact of sampling strong temperature gradients with poorer
spatial resolution.  The fact that we generally find 2--$T$ models to
yield better fits and higher abundances only for groups where our
innermost bin covers a large fraction of $D_{25}$, would seem to add
support for this interpretation.  Indeed, as discussed in Paper~II,
for spectra covering the full region inside $0.1r_{500}$ (a region
much larger than that covered by the central bin in all groups), we
find that two-phase models are in most cases statistically required.
In addition, we note that the introduction of a $\Gamma \approx 1.7$
power-law to account for unresolved X-ray binaries in the central
early-type also mitigates the need for a 2-$T$ model in some cases.

It is finally worth noting that any remaining Fe bias would only
strengthen the result that an Fe excess is present in the cores of
most of our groups. Qualitatively, therefore, the only significant
feature potentially induced by the Fe bias would be the off-centre Fe
peak seen in some systems.  However, some of our groups for which
2--$T$ models provide better fits and significantly enhanced
abundances in the central regions, such as NGC~507, still show
evidence for such Fe and Si peaks, suggesting this feature is
reasonably robust to such a bias.

\subsection{Background estimation}
 
Another important issue is the accuracy of our background subtraction.
This is mainly relevant at large radii, where the emission at all
energies is typically dominated by the background. As mentioned, we
use blank-sky data to extract background spectra, scaled to match the
10--12~keV count rates of our source data for each CCD.  The resulting
medium-to-hard background above $E>4~$~keV seems properly accounted
for in all cases, as there is no statistically significant residual
emission above this energy within the spectral extraction regions
employed for {\em any} data set.  Of potentially more concern is the
soft X-ray background below $E\approx 1$~keV, whose intensity varies
considerably across the sky.
For 10 out of our 15 groups, the {\em ROSAT} All-Sky Survey $R45$
(0.5--0.9~keV) background count rates are within $2\sigma$ ($\la
10$~per~cent deviation) of the analogous value for the relevant
blank-sky background data, lending support to our use of the latter
for background estimation.  While most of our groups are thus not
projected onto regions of significantly enhanced or depleted soft
X-ray background, there are a few exceptions which all show enhanced
background, most notably NGC~5044 and NGC~5846 which both lie close to
the North Polar Spur.

For most of our systems, intragroup emission covers the field of view
($r_{500}$ is outside the field of view in most cases), so we cannot
in general use the outer portions of the observed fields to reliably
assess the level of any residual soft background. Instead, two
independent tests were performed to assess the impact of varying the
assumed background: We either simply rescaled the blank-sky
background, or added a separate background model component in the fits
to the group emission.

In the first test, we repeated all fits for the three outermost radial
bins of each group, with the blank-sky background normalization
shifted by $\pm 10$~per~cent.  The fits were performed in the
0.7--2.0~keV band only, to prevent the obvious mismatch resulting at
higher energies from affecting the results.  We found that this in
general induced negligible changes in the best-fitting temperature,
which remained consistent with the value derived for the `unperturbed'
background in all but two cases (when increasing the background for
NGC~4125 and NGC~6338), for which the resulting fits were
statistically unacceptable anyway.  If anything, the best-fitting
value of $T$ is slightly lower in all cases, so there is no indication
that the temperature decline observed in all groups at large radii can
be ascribed to inaccurate background subtraction.  We therefore
conclude that the large-scale behaviour of the temperature profiles in
Fig.~\ref{fig,T} is a robust result.

The resulting abundances are more sensitive to variations in the
assumed background level. For three of the groups (HCG~42, NGC~1407,
and NGC~2300) rescaling the background produced higher abundances in
the outermost bin, inconsistent with the result obtained using our
standard background level.  None of these cases yielded unacceptable
fits, so we conclude that significantly higher values of $Z$ in the
outermost region probed for these systems are allowed, though not
favoured, by the data.  However, at least in the case of HCG~42, there
is good reason to believe that the existing abundance measurement is
fairly reliable, as it compares well to the result derived
independently from {\em XMM} data \citep{fino07}. Similar agreement is
found for NGC~2300 when compared to earlier {\em ROSAT} and {\em ASCA}
results (see \citealt{rasm06}).  As an additional test, we also tried
varying the background normalization by $\pm 20$~per~cent for all
groups. In the four cases where the resulting fits did not become
either significantly worse or statistically unacceptable, the derived
$T$ and $Z$ were consistent with the results obtained using the
unperturbed background.

As a second test, carried out for the five groups with significantly
enhanced soft background, we included a spectral model to account for
this excess background, and repeated all fits outside the cool core
with this model added to that used to describe the group emission.
Following \citet{vikh05} and \citet{osul07}, a fixed $T=0.18$~keV,
$Z=$~Z$_\odot$ MEKAL background model was assumed, with only its
normalization allowed to vary. Only for NGC~5846, which -- along with
NGC~5044 -- shows the highest background excess, did we find
significant changes in fitted parameters, with this (improved) fit
yielding lower Fe and Si abundances, which we have consequently
adopted. For NGC~5044, no changes were seen, possibly because of the
relatively high signal-to-noise ratio in the outermost bins for this
group.

We confirmed that soft background excess is not an issue for the
remainder of the sample. For example, for relatively compact
low--surface-brightness sources such as HCG~42 and NGC~4125, there is
no statistically significant residual emission in the 0.7--2~keV band
in a 2~arcmin wide annulus extending to the edge of the field of view.
For the adopted soft background model, only 15~per~cent of the total
0.5--0.9~keV flux (used to determine the presence of a soft excess)
falls at energies above 0.7~keV.  The fact that energies below 0.7~keV
are excluded in spectral fitting should therefore also reduce the
problem.  Finally, we do not see any systematic low-energy ($\sim
0.7$--1~keV) residuals for any of the spectra.  This suggests that
systematic uncertainties in our background subtraction is not a major
issue at low energies either.

\section{Combining the results}\label{sec,combine}

A radial profile of some quantity may not always provide a useful
description for an individual system due to departures from spherical
symmetry. The average of many such profiles derived for different
groups should be much less affected though.  Since our radial
measurements for individual groups are statistically independent (as
they are based on projected profiles), it is possible to explore
general trends in the radial distributions of $T$ and $Z$ by
superposing and averaging the profiles derived for individual groups.
With the robustness of our results well established, we can therefore
proceed to discuss these in a statistical framework on the basis of
correlation tests and orthogonal regression analyses \citep{isob90}
performed on the combined measurements.
Throughout the section, we have used the Kendall rank-order
correlation coefficient $\tau$ (in the range $[-1; +1]$, with $+1$ for
a strong, positive correlation), and the associated significance
$\sigma_K$ of $\tau$ being non-zero, to quantify the strength and
nature of any linear relationship between various quantities.

\subsection{Temperature profiles}\label{sec,temp2}
 
In order to examine the overall radial behaviour of $T$ for our
sample, we plot in Fig.~\ref{fig,Tnorm} all temperature measurements,
normalized to the mean temperature $\langle T \rangle$ of each
group.  In the plot, we have subdivided the data points according to
whether they are inside or outside the $T$-peak located at $r(T_{\rm
  max})$, resulting in 134 and 58 measurements, respectively.  For
NGC~4125, which has a monotonically decreasing $T(r)$, we have grouped
the data points inside $D_{25}$ with those at $r<r(T_{\rm max})$.  As
discussed, temperature variations clearly exhibit a very similar
overall behaviour.  Similar to the case for clusters (e.g.\
\citealt{vikh05,prat07}), and as discussed quantitatively below, there
is a clear indication that the intrinsic scatter in $T/\langle T
\rangle$ is larger in the central regions, inwards of the temperature
peak. This is not surprising, as non-gravitational processes other
than radiative cooling could be important in this region, thus helping
to break any self-similarity otherwise induced by gravity.

\begin{figure*} 
\begin{center} 
\mbox{\hspace{-7mm} 
 \includegraphics[width=130mm]{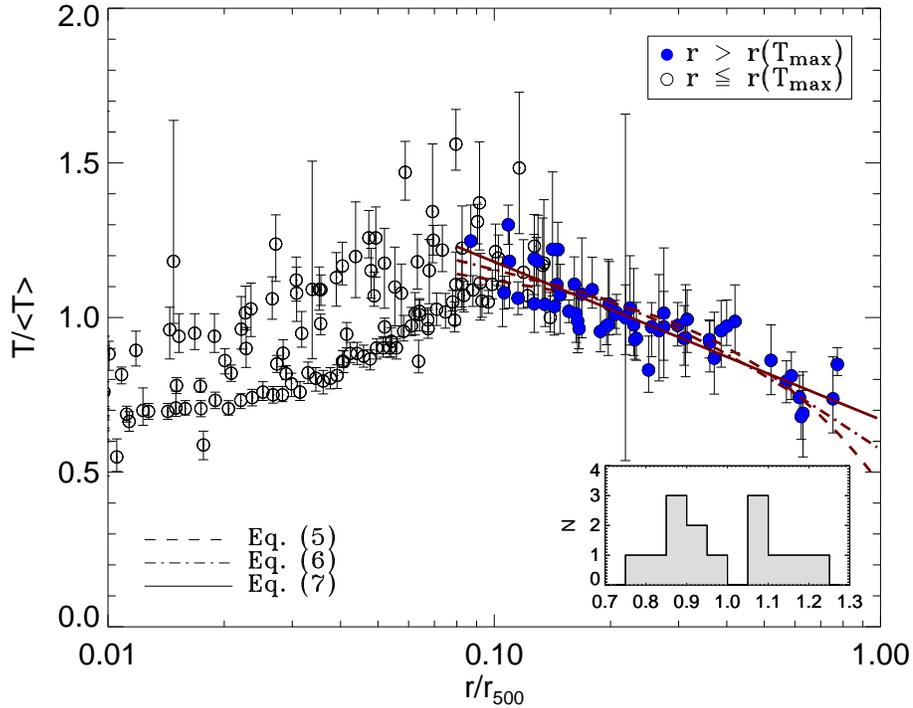}} 
\caption{Temperature profiles of all groups, normalized to the mean
  temperature $\langle T\rangle$ of each group. Filled circles
  represent data points outside the position $r(T_{\rm max})$ of the
  temperature peak (i.e.\ outside the cool core), empty circles those
  inside. Lines show results of regression fits to the data, assuming
  various functional forms of $T/\langle T\rangle $ vs.\ $r$ as
  described in equations~(\ref{eq,T1})--(\ref{eq,T}). Inset shows the
  histogram of $T/\langle T\rangle$ at $r\approx 0.03r_{500}$ for the
  14 cool-core groups.}
\label{fig,Tnorm} 
\end{center} 
\end{figure*}

At first glance, Figure~\ref{fig,Tnorm} appears to be suggestive of a
bimodality in $T(r)$ in the core regions, but this could be largely
driven by NGC~5044, which is better sampled than the other groups and
displays lower central values of $T/\langle T\rangle$ than the
majority. To investigate this in more detail, we tested for any
evidence of bimodality in $T/\langle T\rangle$ by extracting the value
for each group around a specific overdensity radius inside the cool
core, excluding the non-cool core system NGC\,4125. For $r\approx 0.01
r_{500}$, the smallest radius shown in Fig.~\ref{fig,Tnorm}, a
Kolmogorov-Smirnov test indicates a 65~per~cent probability that the
resulting values have been drawn from a single Gaussian distribution
with the given mean and variance. At $0.03r_{500}$, however, there is
some evidence for bimodality (see inset in Fig.~\ref{fig,Tnorm}), with
the corresponding probability being only 12~per~cent, but increasing
again to 45 and 91~per~cent at 0.05 and $0.07r_{500}$, respectively.
Intriguingly, we conclude that there is some evidence for a bimodal
$T/\langle T\rangle$ distribution around $r\approx 0.03r_{500}$, but
the result is marginal in the sense that a unimodal Gaussian
distribution cannot be rejected at 90~per~cent confidence, in part due
to the relatively small sample size. There is also no strong
indications of bimodality at significantly smaller or larger radii.

In order to investigate the potential role of AGN feedback in
establishing this tentative bimodality, we searched for a central
radio source in the brightest group galaxy following the procedure
outlined in \citet{cros05}. Such a source was clearly identified in 11
of the groups (with the off-centre source in NGC\,4125 as an
additional candidate), and with HCG\,42, MKW\,4, and NGC\,4325 as the
only clear exceptions. For these 11 groups, testing for a linear
correlation between $T/\langle T\rangle$ at $r\approx 0.03r_{500}$ and
the 1.4~GHz power of this central radio source yields a value of
Kendall's $\tau = +0.31$ and an associated significance of $\sigma_K =
+1.3$. Thus, there is marginal statistical evidence for a correlation,
but this is largely driven by the two most powerful sources (in
NGC\,383 and NGC\,741), without which any correlation is absent
($\sigma_K=+0.4$). In conclusion, there appears to be no clear
connection between this potential $T(r)$ bimodality and the current
radio output of a central AGN. A larger sample would be useful for a
more decisive test of the presence of a bimodality, and for addressing
its possible origin.

It is also worth noting that we are probing the cores of these groups
down to much smaller fractions of $r_{500}$ than in typical cluster
studies. Interestingly, excluding the non--cool core system NGC~4125,
the ratio $T_c/\langle T\rangle$ of the temperature $T_c$ in the
innermost radial bin to the mean temperature $\langle T\rangle$ shows
a mean and standard deviation of 0.58 and 0.14, respectively (0.61 and
0.17 if NGC~4125 is included). This is higher than the typical
cool-core cluster value of $T_c/\langle T\rangle \approx 0.4$ found,
for example, for the projected $T$-profiles of \citet{vikh05} or the
deprojected ones of \citet{piff05} and \citet{sand06}. It may reflect
the fact that the temperature profiles of our groups flatten off at
small radii (inside $r\approx 0.01r_{500}$), as suggested in
Section~\ref{sec,temp}, in contrast to what is seen in clusters.
Fig.~\ref{fig,projct} indicates that deprojecting our profiles would
not change this conclusion, as already suggested by the internal
consistency among projected and deprojected cluster results.  

An orthogonal regression fit to the temperature profiles inside the
cool core -- excluding the innermost $r<0.01r_{500}$ because of the
putative flattening in this region -- yields
\begin{equation}
  \mbox{log }T/\langle T\rangle = 
  (+0.21\pm 0.02)\mbox{ log}(r/r_{500 }) + (0.28\pm 0.03).
  \label{eq,Tcc}
\end{equation}
The derived logarithmic slope of the profile in the core of $\approx
0.2$ is in excellent agreement with the value of $0.23\pm 0.04$ seen
for groups in the radio-quiet (1.4~GHz radio power of central
early-type $L<10^{22}$~W~Hz$^{-1}$) subsample of \citet{jeth07}. It
is, however, somewhat flatter than the slope of $0.34\pm 0.04$ seen
for their radio-loud subsample, or that of $\approx 0.4$ found for
clusters by \citet{sand06}.

Hence, not only do the temperature profiles appear to flatten off in
the very core of these groups, their rise with radius is also somewhat
shallower than in clusters.  We investigate these results from a
slightly different perspective in Figure~\ref{fig,TcTmax}, plotting
the ratio of $T_c$ to the peak temperature $T_{\rm max}$ for each
group. While this ratio is subject to considerable scatter at a given
mean system temperature $\langle T\rangle$, there is a weak indication
($\tau=-0.43$, $\sigma_K=-1.3$) of a negative correlation between the
quantities plotted (the low-lying outlier in Fig.~\ref{fig,TcTmax} is
NGC~2300; if one were to exclude this data point, the significance of
the correlation would increase to $2.2\sigma$), thus confirming the
notion that the cool core is less pronounced in groups than in
clusters.  The implication seems to be that, relative to the maximum
temperature of the system, cooling in clusters can generally progress
further than in groups, and this group--cluster discrepancy becomes
increasingly pronounced towards lower mean system temperatures
$\langle T \rangle$.

\begin{figure} 
\begin{center} 
\mbox{\hspace{-2mm} 
 \includegraphics[width=86mm]{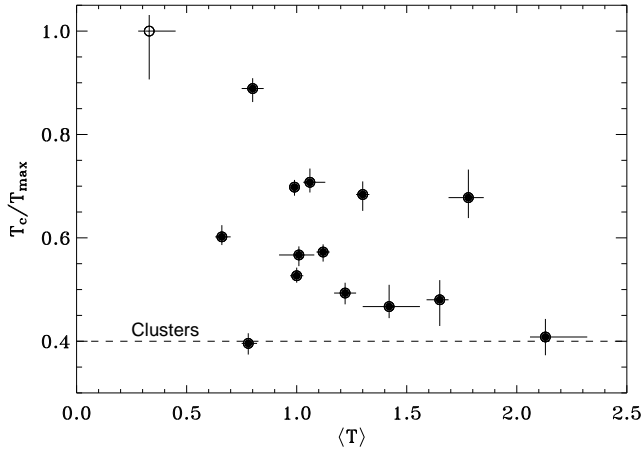}}
\caption{Ratio of core to peak temperature as a function of mean group
  temperature $\langle T \rangle$.  NGC~4125, which has no cool core,
  is shown as an empty circle. Dashed line shows the typical cluster
  value of $T_c/T_{\rm max} \approx 0.4$ (e.g., \citealt{sand06}).}
\label{fig,TcTmax} 
\end{center} 
\end{figure}   

A potentially related issue is the radial extent of the cool cores.
As mentioned in Section~\ref{sec,temp}, the position $r(T_{\rm max})$
of the temperature peak varies slightly among the groups.  In order to
investigate whether this might reflect the properties of the central
galaxy or the group itself, we plot in Fig.~\ref{fig,Tmax} $r(T_{\rm
  max})$ as a function of the size and mass (measured by $D_{25}$ and
$L_K$, respectively) of the central galaxy and of the mass of the
group (measured by $r_{500}$), with $K$-band luminosities $L_K$ taken
from 2MASS \citep{skru06}. Testing for a linear correlation
between $r(T_{\rm max})$ and the three quantities plotted, we find
$\tau = 0.62$ and $\sigma_K = 3.2$, for $r(T_{\rm max})$ vs.\
$r_{500}$.  The corresponding significances for a correlation with
$L_K$ and $D_{25}$ are 2.8 and $2.5\sigma$, respectively.  Although
not a strong result, it suggests that the central temperature
structure of the group is more closely related to the properties of
the group as a whole than to those of the central galaxy (any
correlation with central galaxy properties might naturally arise
because $D_{25}$ and $L_K$ are themselves correlated with $r_{500}$ at
2.5 and $2.7\sigma$ significance, respectively).  Such an
interpretation would be in qualitative agreement with the conclusions
of \citet{hels01} and \citet{hels03} who found that the X-ray
properties of gas surrounding central group galaxies are different
from those of non-central galaxies and, indeed, appear more closely
related to the group than to the central galaxy itself.

\begin{figure*} 
 \includegraphics[width=176mm]{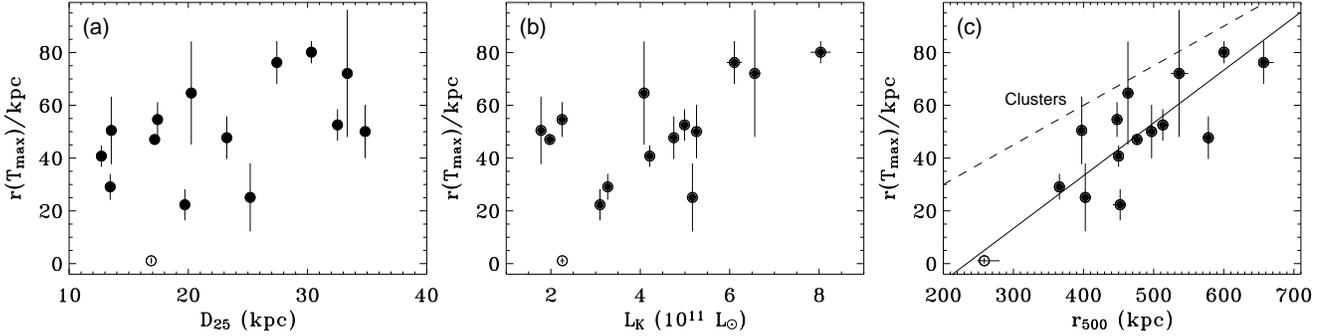} 
 \caption{Radius $r(T_{\rm max})$ of maximum gas temperature as a
   function of (a) $D_{25}$ of the central galaxy, (b)
   extinction-corrected $K$-band luminosity $L_K$ of the central
   galaxy, and (c) $r_{500}$ of the group. As in
   Fig.~\ref{fig,TcTmax}, NGC~4125 is represented by an empty circle.
   The solid line in Fig.~\ref{fig,Tmax}c (with slope = 0.2)
   represents the best fit to the data excluding NGC~4125, while the
   dashed line shows the typical cluster result of $r(T_{\rm
     max})\approx 0.15r_{500}$ (e.g., \citealt{vikh05}).}
\label{fig,Tmax}
\end{figure*} 

Even when excluding NGC~4125, by far the coolest group in our sample,
a fit to the data in Fig.~\ref{fig,Tmax}c gives the relation
\begin{equation} r(T_{\rm max}) = (0.20\pm 0.02) (r_{500}/\mbox{kpc})
  - (46.7\pm 15.1) \mbox{ kpc},
  \label{eq,cc} 
\end{equation} 
which, perhaps coincidentally, would predict the observed absence of a
cool core in NGC~4125. If this relation can indeed be extended to
$r_{500}\la 350$~kpc, the implication would be that systems with
$r_{500} \la 250$~kpc, or, more precisely, with $T\la 0.3$~keV from
equation~(\ref{eq,r500}), would generally not have cool cores. This is
perhaps not surprising, as the central gas properties in systems of
such low temperatures could be strongly affected by galactic feedback.
Measured relative to $r_{500}$, Fig.~\ref{fig,Tmax}c also shows that
while the cool core is generally smaller than in clusters, it
approaches the typical cluster extent at the high end of the
temperature range covered by our sample, with group and cluster
results becoming statistically indistinguishable at $T\ga 2$~keV
according to equations~(\ref{eq,r500}) and (\ref{eq,cc}).

Outside the cool core, the temperature profiles in
Fig.~\ref{fig,Tnorm} clearly decline with $r$, appearing to drop by
almost a factor of two out to $r_{500}$. A correlation test on the
quantities plotted in Fig.~\ref{fig,Tnorm} establishes a strong
anticorrelation beyond $r=r(T_{\rm max})$, with $\tau = -0.70$ and
$\sigma_K = -7.7$. For the data in this region, we attempted three
different parametrizations for this decline of the $T$-profile.  From
a {\em Chandra} analysis of the projected temperature profiles of a
sample of 13 massive, relaxed clusters ($T$-range 1--9~keV),
\citet{vikh05} find a relation $T/\langle T \rangle = -0.76 r/r_{500}
+ 1.22$ outside the cool core (assuming $r_{500} \approx 0.63
r_{180}$).  A similar linear fit for our groups yields
\begin{equation} 
  T/\langle T \rangle = (-0.74\pm 0.09) (r/r_{500}) + (1.20\pm 0.02), 
\label{eq,T1} 
\end{equation} 
for a reduced 
$\chi^2_{\nu}=1.97$.  Taken at face value, this relation, shown as a
dashed line in Fig.~\ref{fig,Tnorm}, is in excellent agreement with
that of the Vikhlinin et~al.\ clusters, suggesting that temperature
variations within groups are very similar to those of clusters at
large radii, contrary to the case in the core.  The resulting fit
quality is poor, however, so a relation of the form of
equation~(\ref{eq,T1}) is not particularly successful in describing
our group data.  The main reason is that such a parametrization cannot
match the group data immediately outside the cool core without
overpredicting observed temperatures at intermediate radii and
underpredicting them further out.
 
As a possible alternative, we note that cosmological hydrodynamical
simulations invoking gravity only have suggested a `universal'
temperature profile for massive ($T\ga 3$~keV) clusters, of the form
$T\propto (1+0.75r/r_{500})^{-1.6}$ \citep{loke02}, assuming
$r_{500}\approx 0.5r_{100}$. This parametrization was shown by
\citet{loke02} to be in good agreement with cluster results outside
the core region, remaining a good description also when including
galactic feedback and radiative cooling in the simulations. Motivated
by this, we fitted a similar relation to our group data, yielding
\begin{equation} 
  T/\langle T \rangle = (1.29\pm 0.03)(1+ 0.75r/r_{500})^{-1.45\pm 0.14} 
\label{eq,T2} 
\end{equation} 
for $\chi^2_{\nu}=1.65$. While resulting in a better fit than
equation~(\ref{eq,T1}), this relation suffers from the same
shortcomings in describing our data, albeit to a lesser degree.
Nominally, the result, represented by a dash-dotted line in
Fig.~\ref{fig,Tnorm}, is broadly consistent with that found for the
simulated clusters of \citet{loke02}, including the normalization
factor of 1.33 found by the latter authors.

Prompted by the appearance of the data in Fig.~\ref{fig,Tnorm}, a
third parametrization of the profiles was also investigated, in which
$T$ is a linear function of $\mbox{log }(r/r_{500})$. The fit result,
\begin{equation}
  T/\langle T\rangle = 
  (-0.51\pm 0.04) \mbox{ log}(r/r_{500}) + (0.67\pm 0.03),  
\label{eq,T} 
\end{equation} 
is statistically better ($\chi^2_{\nu}=1.26$) than the results of
equation~(\ref{eq,T1}) and (\ref{eq,T2}).  The intrinsic scatter about
this relation is negligible, whereas the corresponding scatter around
equation~(\ref{eq,Tcc}) is 12~per~cent, thus confirming the indication
of Fig.~\ref{fig,Tnorm} that intrinsic scatter is larger in the core
regions.  The derived relation of equation~(\ref{eq,T}) is plotted as
a solid line in Fig.~\ref{fig,Tnorm}.
The fact that equations~(\ref{eq,T1}) and (\ref{eq,T2}) are good
descriptions of both observed and simulated clusters, but are less
representative of groups than equation~(\ref{eq,T}), does not
necessarily suggest the presence of intrinsic differences in the
large-scale temperature structure of hot gas in these systems. The
formal agreement with cluster results in terms of best-fitting
parameters clearly indicate that the temperature profiles of cool-core
groups at large radii are similar to those of clusters. We do not
expect this conclusion to be affected by projection effects to an
appreciable degree. Finally, we note that there is no indication that
the possible bimodality in $T/\langle T\rangle$ at $r\approx
0.03r_{500}$ is reflected in significantly different slopes at large
$r$; assuming again a prescription of the form of
equation~(\ref{eq,T}) outside $r(T_{\rm max})$, we find consistent
slopes of $-0.49\pm 0.07$ and $-0.54\pm 0.04$ for the low- and
high-$T/\langle T\rangle$ subsamples seen in the inset of
Fig.~\ref{fig,Tnorm}, respectively.

Summarizing, we find that cool cores in groups are less pronounced
than those of clusters, in terms of both their spatial extent and
relative temperature variation. At large radii, however, the
temperature profiles of our groups decline with radius in consistency
with results for more massive systems.

\subsection{Iron and silicon profiles}\label{sec,iron}
 
In Fig.~\ref{fig,Z_all}, all measurements of $Z_{\rm Fe}$ and $Z_{\rm
  Si}$ have been combined, demonstrating more clearly the central
excess in iron abundance and the general decrease to a value of $\sim
0.1$~Z$_\odot$ at large radii suggested by Fig.~\ref{fig,abund}.  For
$Z_{\rm Fe}$, there appears to be a clear trend with radius outside
the cool core, and a test performed on the data points outside
$r(T_{\rm max})$ confirms the presence of a linear correlation between
log~$Z_{\rm Fe}$ and log~$(r/r_{500})$ at the $5.1\sigma$~level
($\tau=-0.46$). Inside the core, the data are consistent with a
constant Fe abundance ($\sigma_K=0.0$), but this may mask a more
complex behaviour as discussed below.

\begin{figure*} 
\begin{center} 
 \includegraphics[width=150mm]{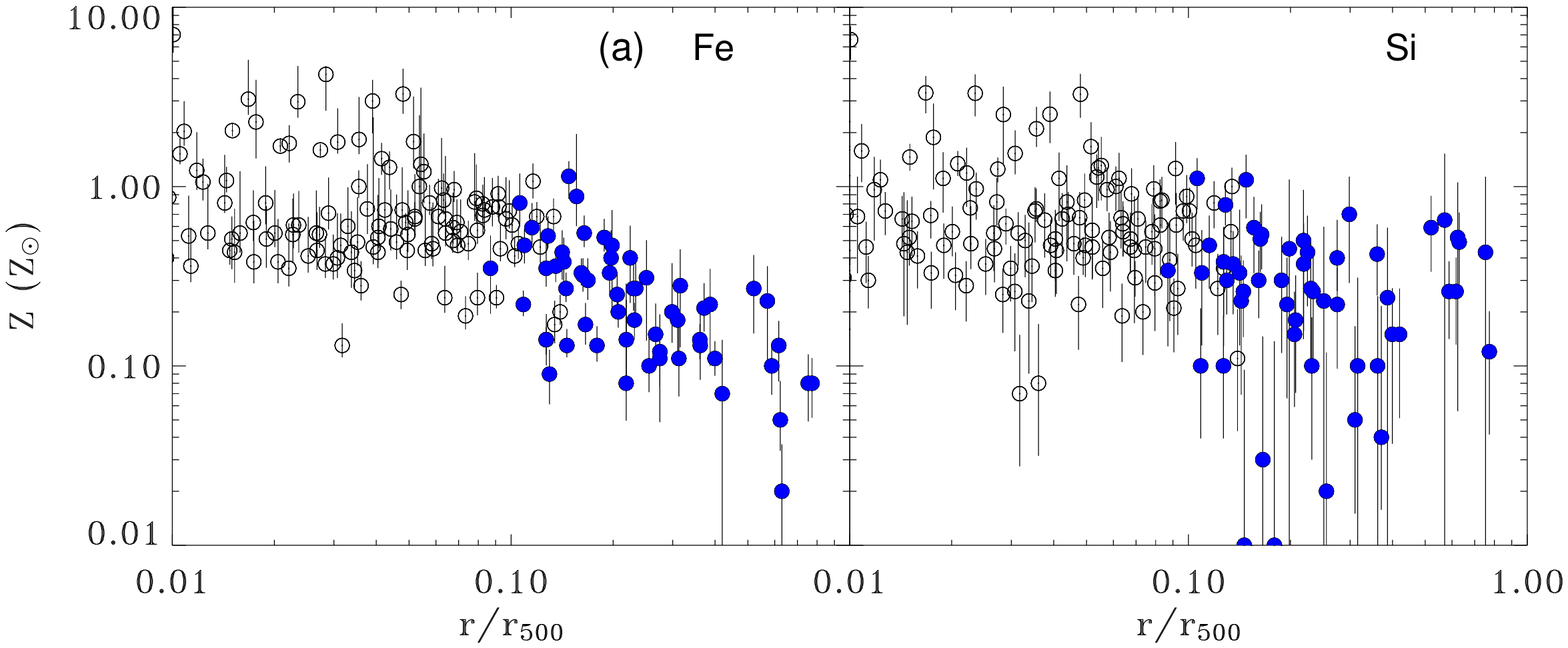} 
 \includegraphics[width=150mm]{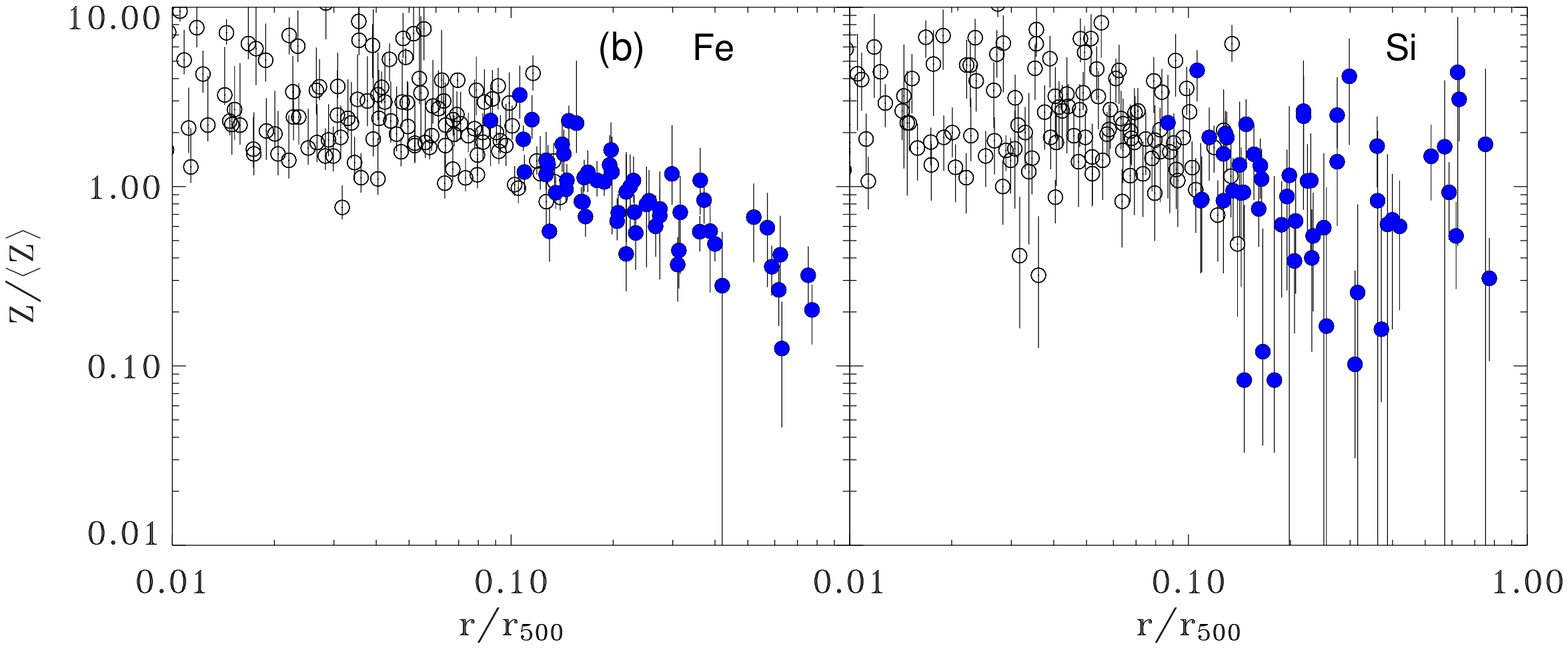} 
 \caption{(a) $Z_{\rm Fe}$ and $Z_{\rm Si}$, all measurements.
   Symbols are as in Fig.~\ref{fig,Tnorm}. (b) The same data
   normalized to $\langle Z\rangle$.}
\label{fig,Z_all} 
\end{center} 
\end{figure*} 

The Si profiles are fairly similar to those of Fe in the group cores,
showing no systematic trend with radius ($\sigma_K=-0.1$).
Si also appears from Fig.~\ref{fig,Z_all} to show less systematic
radial variation outside the core than Fe.  Indeed, outside $r(T_{\rm
  max})$ no significant correlation ($\tau=-0.07$, $\sigma_K=-0.8$) is
found between log~$Z_{\rm Si}$ and log~$(r/r_{500})$, though the value
of $\tau$ suggests a mild overall decline with radius. The situation
is qualitatively similar to that seen for clusters, where Si generally
shows a more uniform distribution than Fe \citep{fino00}. As is
evident from Fig.~\ref{fig,abund}, $Z_{\rm Si}$ actually rises with
radius in the outskirts for some of the groups, a fact which clearly
contributes to obscuring any statistical evidence for a simple linear
behaviour. A similar feature has been noted in some relatively cool
clusters but is seemingly absent in hotter systems
\citep{fino00,fino01a}.

In addition to the considerable statistical errors on the derived
abundance data, particularly for Si at large radii, there is clearly
also substantial intrinsic scatter among the groups at any given
radius. Coupled with the lack of a clear radial correlation for Si, we
have consequently refrained from fitting regression lines to the raw
data in Fig.~\ref{fig,Z_all}a.  In an attempt to suppress these
intrinsic variations, we also plot in Fig.~\ref{fig,Z_all}b the
abundance data normalized to the mean value $\langle Z \rangle$. As
suggested by the figure, the resulting correlation is slightly
stronger for $Z_{\rm Fe}$ ($\tau=-0.61$, $\sigma_K =-6.8$) but is
still not very significant for Si ($\sigma_K=-1.2$). The latter result
is perhaps not surprising, given that the normalization factor, i.e.\
the mean abundance $\langle Z \rangle$, has been derived assuming all
abundances fixed relative to solar, so the fitted value of $\langle Z
\rangle$ will be driven by the abundance as estimated from the
prominent Fe lines.

\subsection{The silicon-to-iron ratio $Z_{\rm Si}/Z_{\rm Fe}$}   

The strong radial decline of $Z_{\rm Fe}$ and the hint of a more
gentle decline of $Z_{\rm Si}$ are also reflected in an increase in
$Z_{\rm Si}/Z_{\rm Fe}$ at large radii, as illustrated in
Fig.~\ref{fig,ratio_all}.  This figure further suggests the presence
of a dichotomy in the radial distribution of $Z_{\rm Si}/Z_{\rm Fe}$,
with this quantity being roughly constant inside $r(T_{\rm max})$ but
rising further out.  A correlation test confirms that within the cool
core there is only marginal evidence for a systematic variation in
this quantity ($\tau=0.08$, $\sigma_K=1.3$), suggesting a mild radial
decrease, with the data showing a mean and $1\sigma$ dispersion pf
$0.91\pm 0.29$~Z$_{{\rm Si},\odot}$/Z$_{{\rm Fe},\odot}$. This mean
value is thus slightly lower than, but consistent with, the local
(Solar neighbourhood) IMF and SN mixture. For the adopted SN yields,
the value corresponds to a mixture of 70~per~cent core-collapse SN and
30~per~cent SN~Ia by number, in excellent agreement with recent
results for the cores of clusters (see \citealt{depl07} and references
therein), and consistent with the observed proportion of SN types at
low redshift ($z\la 0.5$) determined from the {\em Hubble} and {\em
  Chandra} Deep Fields \citep{dahl04}. For the adopted SN yields, this
implies that SN~Ia are the dominant Fe contributor in group cores,
being responsible for $79\pm 13$~per~cent of the Fe in this region.
This agrees not only with results for cluster cores, as mentioned, but
is also in accord with the corresponding fraction for the hot gas in
elliptical galaxies of $66\pm 11$~per~cent \citep{hump06}.

Outside $r(T_{\rm max})$, the correlation test confirms that $Z_{\rm
  Si}/Z_{\rm Fe}$ rises with radius ($\tau=0.21$, $\sigma_K=2.3$), but
the trend is probably suppressed due to large statistical scatter.
We note that many of the data points with very low $Z_{\rm Si}/Z_{\rm
  Fe}$ have $Z_{\rm Si}\approx 0$ rather than high $Z_{\rm Fe}$, but
these have large fractional errors and most are consistent with the
general trend.  Similar to the case for clusters and as further
illustrated in Fig.~\ref{fig,ratio_all}b, the high central Fe
abundance seen in most of our systems can thus be ascribed
predominantly to SN~Ia, with an increasing SN~II contribution towards
larger radii and lower Fe abundances.  We also note that the global
mean and standard deviation of $Z_{\rm Si}/Z_{\rm Fe}$, derived from
those data points for which the fractional errors (at 90~per~cent
confidence) on both $Z_{\rm Si}$ and $Z_{\rm Fe}$ are less than
100~per~cent, are $\approx 1.3$ and 0.8 in solar units, respectively,
in agreement with the (radially constant) ratio of $1.4\pm 0.2$ found
for the $T<6$~keV subsample of \citet{tamu04}.
The fact that our groups do {\em not} show a radially constant ratio
is reflected in a large standard deviation around the global mean.

\begin{figure} 
\hspace{-5mm}
 \includegraphics[width=90mm]{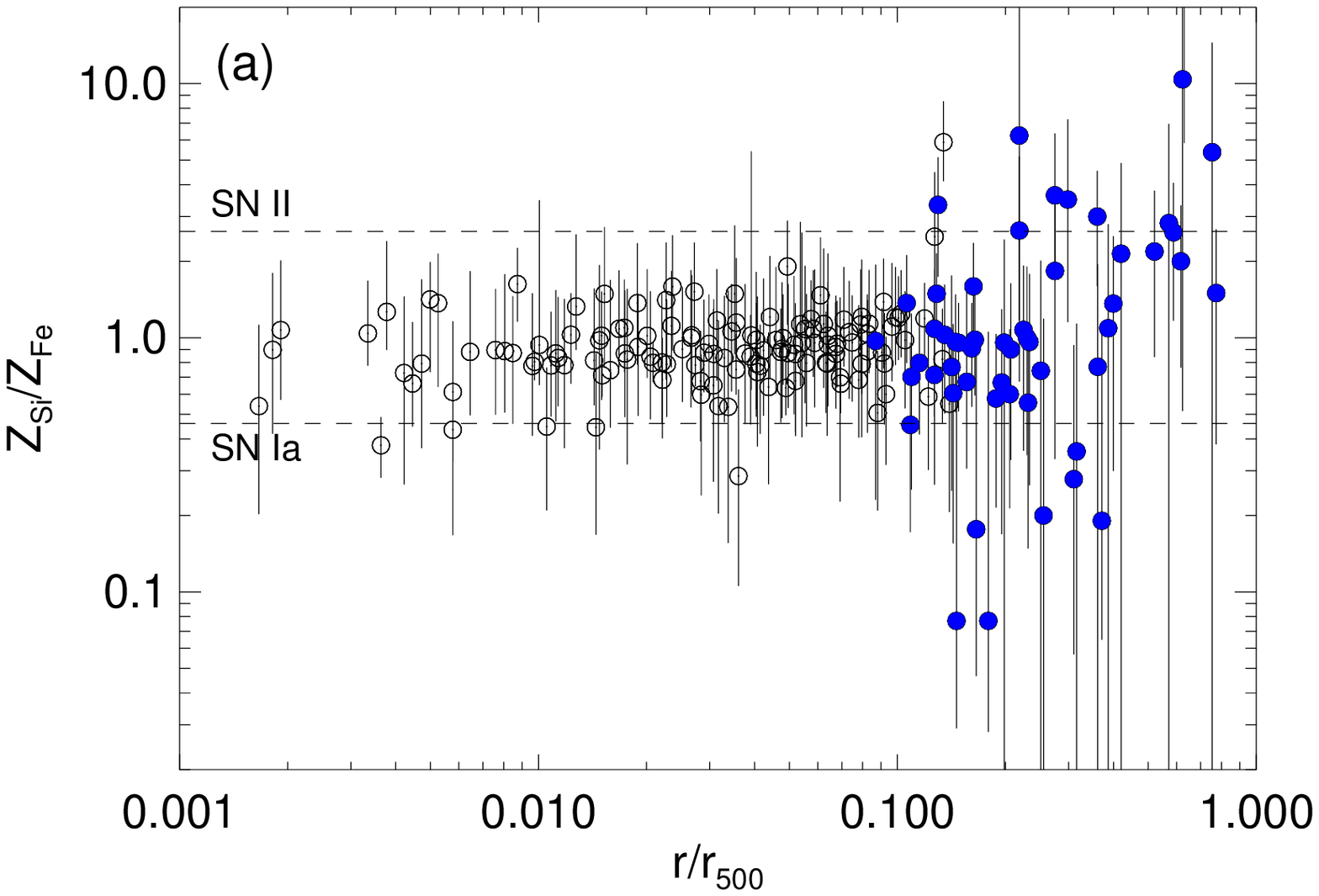}
  
\hspace{-5mm}
\includegraphics[width=90mm]{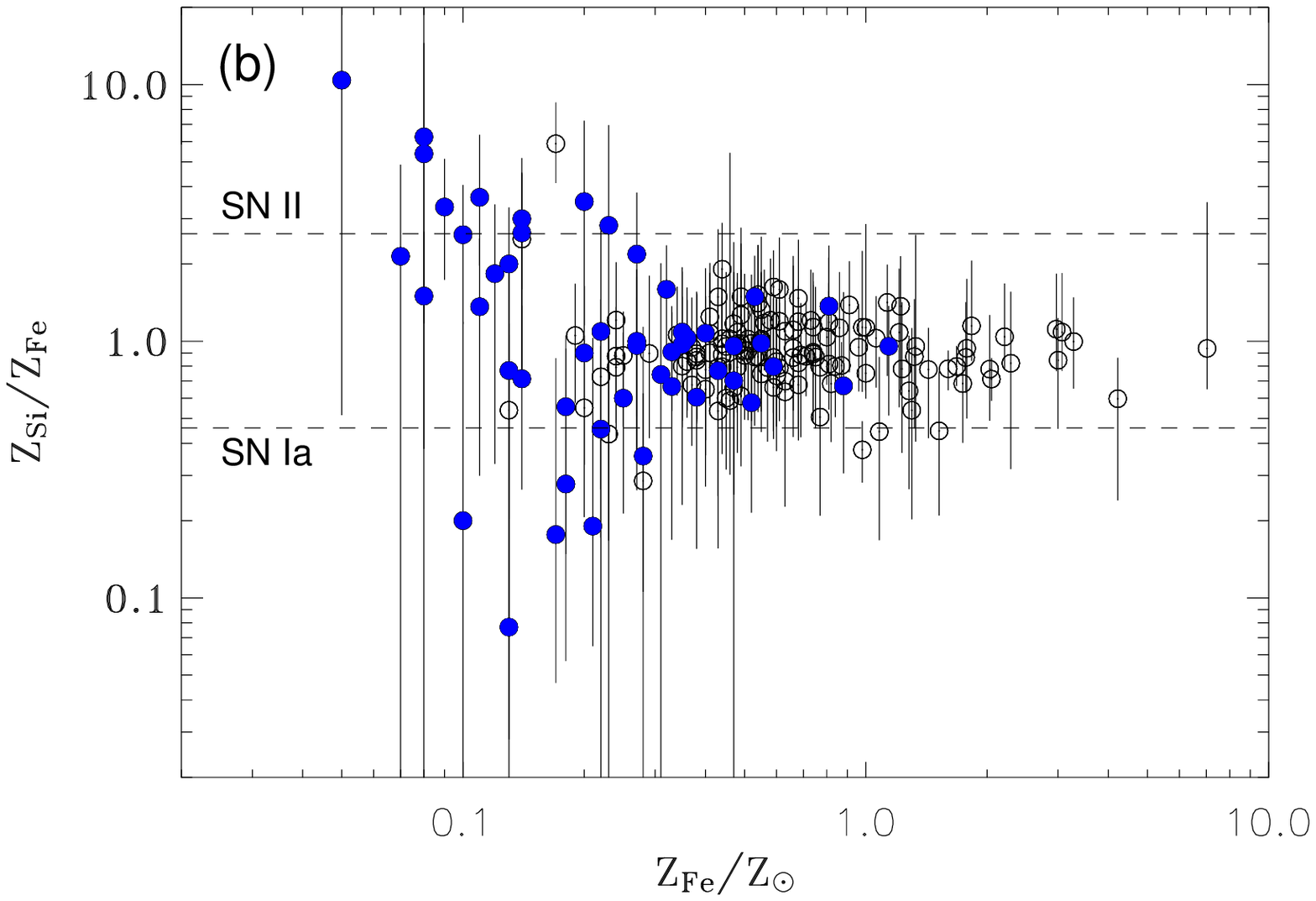} 
 \caption{$Z_{\rm Si}/Z_{\rm Fe}$, all measurements, in units of the
   Solar ratio and as a function of (a) radius and (b) $Z_{\rm Fe}$
   (with error bars on $Z_{\rm Fe}$ omitted for clarity).  Symbols are
   as in Fig.~\ref{fig,Tnorm}. Dashed lines mark the ratios expected
   from pure SN~Ia and SN~II enrichment, as in Fig.~\ref{fig,ratio}.}
\label{fig,ratio_all} 
\end{figure}

\subsection{Binning the profiles} 
 
To better investigate radial trends and further suppress intrinsic
scatter among the groups, we plot profiles of all data points in
Fig.~\ref{fig,norm_bin}, accumulated in radial bins of 20
measurements. In this representation, the temperature profiles are
clearly seen in Fig.~\ref{fig,norm_bin}a to peak at a radius $r\approx
0.1r_{500}$, while flattening out in the group cores. We re-iterate
that this central flattening is not observed in clusters, even in
cases where the cluster cores are well resolved on scales below
$0.01r_{500}$ (e.g., \citealt{sand06}).

For the Fe abundance, Fig.~\ref{fig,norm_bin}b reveals roughly solar
abundances in the group cores, consistent with the results found for
X-ray bright ellipticals by \citet{hump06}, five of which constitute
the central group elliptical in our sample. The figure also confirms
the trends discussed above, indicating that outside the core, $Z_{\rm
  Fe}$ continuously declines out to the largest measured radii.  The
binned results further enable a straightforward comparison to the
corresponding Fe profile found for cool-core clusters by
\citet{degr04}. This reveals similar values of Fe enrichment within
the central abundance excess in groups and clusters ($Z_{\rm
  Fe}\approx 0.8$~Z$_\odot$), with some indication that the excess is
slightly broader in clusters; our binned Fe profile declines from its
central value by a factor of two around $r\approx 0.2r_{500}$, whereas
the corresponding radius for the \citet{degr04} clusters is $r\approx
0.3r_{500}$. The \citet{degr04} clusters display a flattening in the
Fe profile beyond $0.3r_{500}$. contrary to what is observed for our
groups. Note also that the off-centre abundance peaks hinted at in
many of the groups with well-resolved cores, and the considerable
intrinsic scatter these peaks cause in this representation, are
visible for both the binned Fe and Si profiles, even when these are
normalized to the mean abundances.

The binning of the abundance profiles clearly suppresses the
considerable scatter seen in Fig.~\ref{fig,Z_all}, resulting in a much
more well-defined radial behaviour for both Fe and Si. We performed
linear regression fits to these sample-averaged profiles in order to
characterize the reasonably uniform behaviour outside the very group
cores. Excluding the innermost data point in Fig.~\ref{fig,norm_bin}b,
orthogonal regression on the binned Fe profile yields
\begin{equation}
  \mbox{log }Z_{\rm Fe} = (-0.66\pm 0.05)\mbox{ log}(r/r_{500 })-(1.00\pm 0.06)
\label{eq,ZFe} 
\end{equation} 
relative to the corresponding solar abundances.
The Fe abundance at $r_{500}$ suggested by this relation of $Z_{\rm
  Fe}\simeq 0.1$~Z$_\odot$ is lower than the typical value of $\sim
0.2$~Z$_\odot$ seen in the outskirts (at $r \approx r_{500}$) of
clusters (e.g., \citealt{fino00}; \citealt{degr01}; \citealt{tamu04}).
This is particularly true in light of the fact that we have adopted
the abundance table of \citet{grev98}, for which the Solar Fe
abundance is only 68~per~cent of the \citet{ande89} value typically
used in previous cluster studies. The Fe abundance in those previous
studies must therefore be multiplied by 1.48 for a direct comparison
to our results (the Si abundance, on the other hand, remains
unaffected to within 1~per~cent by this change in abundance table).

Fig.\ref{fig,norm_bin}c suggests that $Z_{\rm Si}$ also declines
outside the cool core, despite the fact that there is only marginal
statistical evidence for a radial decline of $Z_{\rm Si}$ outside
$r(T_{\rm max})$ in the unbinned data.  As already indicated by the
dichotomy in the Si/Fe ratio (Fig.~\ref{fig,ratio_all}), the reason
for this apparent inconsistency is that while Si does decline
immediately outside the cool core, it flattens out or even rises at
the largest radii probed, thus suppressing any statistical evidence
for a uniformly declining profile outside the core. Beyond
$0.4r_{500}$, roughly corresponding to the range covered by the
outermost data point in Fig.~\ref{fig,norm_bin}c, the unbinned Si data
show no clear correlation with $r$ ($\sigma_K = -0.7$).  To capture
this behaviour, we performed a regression fit to all but the innermost
data point in Fig.\ref{fig,norm_bin}c, and assumed $Z_{\rm Si}(r)$ to
remain constant beyond the outermost data point at $r\approx
0.5r_{500}$, yielding
\begin{equation}
  \mbox{log }Z_{\rm Si} = 
  \cases{m\mbox{ log}(r/r_{500 })+c ,\,\,\,\, r\leq 0.5r_{500}  \cr 
    0.25\pm 0.05\mbox{~Z$_\odot$}\mbox{\hspace{4mm}},\,\,\,\, r> 0.5r_{500}} ,
\label{eq,ZSi} 
\end{equation}
with $m=-0.44\pm 0.07$ and $c=-0.74\pm 0.06$. The value of
0.25~Z$_\odot$ at $r> 0.5r_{500}$ was chosen to ensure continuity,
with its errors based on the fractional errors on the outermost data
point. 
An Si abundance of $\sim$~0.2--0.25~Z$_\odot$ at $r_{500}$ is also in
good agreement with earlier {\em ASCA} results for groups
\citep{fino02}, whereas clusters tend to show slightly higher values
of $\sim 0.3$--0.5~Z$_\odot$ \citep{fino00,fino01a} at large radii.
This suggests a higher content of {\em both} Fe and Si in the
outskirts of more massive systems.  However, a direct comparison is
somewhat impeded by the large statistical uncertainties on $Z_{\rm
  Si}$ in groups, as well as the fact that $Z_{\rm Si}(r)$ does not
show a uniform behaviour at large $r$ across our sample.

For the $Z_{\rm Si}/Z_{\rm Fe}$ ratio, a subsolar value is seen in the
innermost bin in Fig.~\ref{fig,norm_bin}f, with the ratio rising to
$\approx 1$ within $0.1r_{500}$ and further to $\ga 2$ well outside
the central galaxy. As discussed in Section~\ref{sec,syst}, we believe
that the result for the outermost bin is not seriously affected by
uncertainties in our background subtraction.  The result demonstrates
more clearly what was already indicated by Fig.~\ref{fig,ratio_all},
that SN~Ia provide a dominant, but not exclusive, contribution to the
enrichment within the cool core and hence inside the optical extent of
the central galaxy. Well outside the group core the dominant
contribution comes from SN~II, and at large radii the $Z_{\rm
  Si}/Z_{\rm Fe}$ ratio rises to become consistent with the value of
$\approx 2.6$ expected for pure SN~II enrichment. Corrected for
differences in adopted abundance tables, the value of the outermost
bin in Fig.~\ref{fig,norm_bin}f is at least as high those of the
\citet{fino00} clusters. Our results, therefore, do not support the
claim of the latter authors that groups on average have lower Si/Fe
ratios at large radii than clusters.  The results further indicate
that, on average, both SN~Ia and SN~II enrichment is required at all
radii within at least $r\approx 0.5r_{500}$. While this relies on the
assumption that derived abundances in the central region are not
seriously affected by the Fe and Si biases, similar central $Z_{\rm
  Si}/Z_{\rm Fe}$ ratios of $\ga 1$ have also been found in {\em ASCA}
and {\em XMM} observations of hotter clusters \citep{fino00,tamu04},
although it should be mentioned that the cores of these clusters are
generally less well resolved than those of our groups.

\begin{figure*} 
\hspace{-3mm}
 \includegraphics[width=176mm]{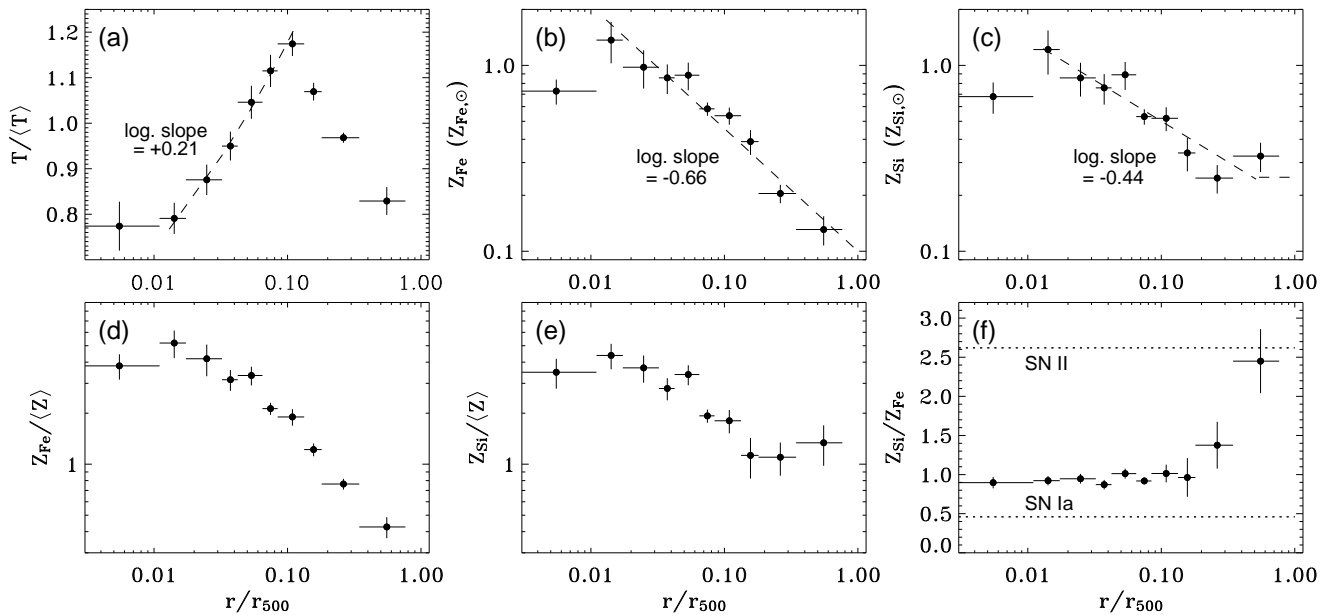} 
 \caption{Profiles for all groups, binned into radial bins of $N=20$
   data points. Errors represent the standard deviation in each bin
   divided by $\sqrt{N}$. Dashed lines in (a)--(c) mark the
   best-fitting regression lines across the radial ranges shown, i.e.\
   equations~(\ref{eq,Tcc}), (\ref{eq,ZFe}), and (\ref{eq,ZSi}).}
\label{fig,norm_bin} 
\end{figure*}

Summarizing, we find that the distributions of both Fe and Si in the
unbinned data (Fig.~\ref{fig,Z_all}) appear roughly uniform in the
group cores, but the averaged profiles clearly confirm the evidence
for off-centre peaks in some of the groups. The enrichment in the
cores contain a significant contribution from SN~Ia, probably hosted
by the central, bright galaxy, an issue to which we will return in
Paper~II. On average, SN~II dominate by {\em number} at all radii, in
agreement with the results obtained by \citet{fino99} for a sample of
three groups, and can on their own explain the observed abundances
well outside the group core.  We believe this latter conclusion to be
reasonably robust with respect to any remaining background issues.  A
high Si/Fe ratio consistent with, or even exceeding, the expectation
from our adopted SN~II yields is seen in the outermost bin of {\em
  all} of our groups, including groups such as NGC~383, HCG~62, and
NGC~5044, for which the data feature high signal-to-noise ratios even
at large radii. Support for this result comes from the work of
\citet{fino00}, in which a rise in $Z_{\rm Si}/Z_{\rm Fe}$ to
supersolar values at large radii is also seen for the groups in their
sample.  The trend of Si/Fe increasing with radius has also been
reproduced in cosmological simulations in which the galactic feedback
implementation favours the ejection of SN~II products over those of
SN~Ia \citep{rome06}.

We note that the interpretation that SN~II alone can explain the
observed abundance pattern at large radii is reasonably robust to the
choice of theoretical SN~II yields available in the literature. For
example, for a standard Salpeter IMF, none of the SN~II model yields
considered by \citet{gibs97} would predict $Z_{\rm Si}/Z_{\rm Fe} >
2.6$. Hence our conclusion regarding the SN~II contribution is only
compromised if invoking a more `top-heavy' IMF, with $x<1.0$ in
equation~(\ref{eq,SN}). In connection with this, it is worth pointing
out, on the basis of the outermost three data points of
Fig.~\ref{fig,norm_bin}f, that the slope of $Z(r)$ for at least one of
the two elements must change beyond $r_{500}$ if the observed Si/Fe
ratio is to remain consistent with the IMF and SN model yields adopted
here. Deep, high--S/N X-ray observations of group outskirts could shed
light on this issue and thus help constrain theoretical SN yields, but
such a study could well be beyond the capabilities of current X-ray
telescopes for all reasonable exposure times. With their higher X-ray
surface brightness at large radii, cool to intermediate-temperature
clusters ($T \approx 2-4$~keV) remain more attractive targets for this
purpose, also because they show prominent emission lines from a wider
range of relevant elements (e.g., \citealt{fino00,fino01a}).

\section{Summary and conclusions}\label{sec,conclude} 

We have conducted a homogeneous analysis of a sample of 15 X-ray
bright galaxy groups observed by {\em Chandra}, selected for their
relatively undisturbed X-ray morphology and good photon statistics.
Correlation tests and regression fits have been performed on the
combined radial temperature and abundance profiles and derived
quantities, both to investigate the nature of any relationship between
various quantities, and to enable straightforward comparison to
results of numerical simulations of group formation and evolution.

Our derived radial temperature profiles show remarkable similarity
when scaled to the mean system temperature $\langle T\rangle$ derived
outside the central region, and reveal that all but one group show
clear evidence for a cool core.  We find that these cores are smaller
than those seen in clusters, both in absolute physical extent and
relative to a fixed overdensity radius, and that their sizes appear
more closely related to the properties of the group as a whole than to
those of the central early-type galaxy present in all our groups. The
absolute and relative cool core sizes approach those of clusters at
the high--temperature end of our sample.  The temperature profiles in
the core are also shallower than those seen in clusters, equivalent to
a less pronounced temperature decline in the core relative to the peak
temperature displayed by the system. This may indicate that radiative
cooling can generally progress further in more massive systems.  The
flatter group profiles inwards of the temperature peak relative to
those of more massive systems certainly point to clear differences in
the central ICM heating history among systems of varying mass. We find
tentative evidence for a bimodality in $T/\langle T\rangle$ inside the
cool core, but this is confined to a narrow interval in radius and is
not mirrored by differing slopes in the temperature profiles at large
radii, which we find to be similar to those found for both observed
and simulated clusters.

The Fe and Si abundance profiles show a number of distinct features
common to most of the groups. Similar to the case for cool-core
clusters, a central enhancement in Fe abundance is clearly seen in all
but two groups, one of which is NGC~4125, the only non--cool core
group in our sample. This Fe excess can extend well beyond the optical
extent of the central early-type galaxy present in all groups, and is
in many cases accompanied by a similar feature in the Si distribution.
In many of the groups with a well-resolved core, evidence for an
off-centre abundance peak is also observed inside the central galaxy,
with a decline in abundance at the very centre, similar to results for
some well-resolved cluster cores. Outside the central excess, the Fe
profiles generally show a steep decline, approaching a value of
$Z_{\rm Fe}\approx 0.1$~solar at $r_{500}$.  This is lower than the
corresponding typical cluster value by a factor of two, suggesting
fundamental differences in enrichment history for the outskirts of
these low-mass systems compared to more massive clusters. Si shows
similar overall features but declines less steeply with radius, and
flattens out or even rises again at large radii for the majority of
the groups. On average, Si reaches a value of $Z_{\rm Si}\approx
0.25-0.3$~solar at $r_{500}$, again slightly below typical cluster
values.

For the adopted SN yields, the ratio of Si to Fe abundance in the
groups reveals a clear predominance of SN~Ia enrichment in the central
regions, with $79\pm 13$~per~cent of the Fe originating in SN~Ia. This
is in line with results for both hot gas in elliptical galaxies and
the cores of massive clusters, suggesting a very similar enrichment
istory for the central regions in such systems across three orders of
magnitude in total system mass. The SN~II contribution in our groups
increases with radius, however, and at the largest radii probed, or,
equivalently, at the lowest Fe abundances, the Si/Fe ratio is close
to, and nearly always consistent with, the expectation for pure SN~II
enrichment in all groups. We find no indication that Si/Fe ratios at
large radii are significantly different from the corresponding values
in clusters.  We note, however, that the existence of two separate
SN~Ia populations, as suggested by \citet{mann06}, could complicate
the interpretation of these results.

The abundance distribution in the intragroup medium is expected to
be tightly linked to the star formation history in groups.
In a forthcoming companion paper, we decompose the derived abundance
profiles into contributions from the two major supernova types to explore this
issue, along with the implications of our results for supernova and ICM
energetics, the role of AGN and mergers in redistributing and
reheating the enriched gas, and more generally for the enrichment
history of the groups.

\section*{Acknowledgments} 
 
We thank Stephen Helsdon for his contribution to the early stages of
this project. We are grateful to Alexis Finoguenov and John Osmond for
useful discussions and for providing comparison data for their group
samples prior to publication.  This work has made use of the NASA/IPAC
Extragalactic Database (NED) and the Two Micron All Sky Survey (2MASS)
database. JR acknowledges the support of the European Community
through a Marie Curie Intra-European Fellowship.

\bsp 
 
\label{lastpage} 
 
\end{document}